\documentclass{aastex6}

\fullcollaborationName{The Friends of AASTeX Collaboration}

\usepackage{graphicx}
\usepackage{graphics}
\usepackage[figuresleft]{rotating}
\usepackage{caption3}
\usepackage{amsmath}
\usepackage{mathrsfs}

\begin{document}


\title{Timing Solutions and Single-pulse Properties for Eight Rotating Radio Transients}


\author{B.-Y. Cui\altaffilmark{1}}
\affil{Department of Physics and Astronomy \\
 West Virginia University \\
 Morgantown, WV 26506, USA}

\author{J. Boyles}
\affil{Department of Physics and Astronomy \\
 West Kentucky University \\
 Bowling Green, KY 42101, USA}

\author{M.A. McLaughlin\altaffilmark{1}}
\affil{Department of Physics and Astronomy \\
 West Virginia University \\
 Morgantown, WV 26506, USA}

\and

\author{N.Palliyaguru}
\affil{Physics and Astronomy Department \\
 Texas Tech University \\
 Lubbock, TX 79409-1051}


\altaffiltext{1}{Center for Gravitational Waves and Cosmology, West Virginia University,  Morgantown, WV 26505}

\begin{abstract}

Rotating radio transients (RRATs), loosely defined as objects that are discovered through only their single pulses, are sporadic pulsars that have a wide range of emission properties. For many of them, we must measure their periods and determine timing solutions relying on the timing of their individual pulses, while some of the less sporadic RRATs can be timed by using folding techniques as we do for other pulsars. Here, based on Parkes and Green Bank Telescope (GBT) observations, we introduce our results on eight RRATs including their timing-derived rotation parameters, positions, and dispersion measures (DMs), along with a comparison of the spin-down properties of RRATs and normal pulsars. Using data for 24 RRATs, we find that their period derivatives are generally larger than those of normal pulsars, independent of any intrinsic correlation with period, indicating that RRATs' highly sporadic emission may be associated with intrinsically larger magnetic fields. We carry out Lomb$-$Scargle tests to search for periodicities in RRATs' pulse detection times with long timescales. Periodicities are detected for all targets, with significant candidates of roughly 3.4 hr for PSR J1623$-$0841 and 0.7 hr for PSR J1839$-$0141. We also analyze their single-pulse amplitude distributions, finding that log-normal distributions provide the best fits, as is the case for most pulsars. However, several RRATs exhibit power-law tails, as seen for pulsars emitting giant pulses. This, along with consideration of the selection effects against the detection of weak pulses, imply that RRAT pulses generally represent the tail of a normal intensity distribution.

\end{abstract}




\section{Introduction} \label{sec:intro}

Rotating Radio Transients (RRATs) can generally be defined as pulsars that were originally detectable only through their single pulses and not through standard Fourier techniques \citep{km11}. They were first discovered through single-pulse search reprocessing of the Parkes Multibeam Pulsar Survey data \citep{mll+06,kel+09,km11}. Currently $\sim100$ of these sporadic pulsars are known.\footnote{See {\url{http://astro.phys.wvu.edu/rratalog}}} Long-term monitoring observations show that there is a wide range of spin-down and emission properties for objects originally termed RRATs, with some appearing as normal or nulling pulsars in higher sensitivity or lower frequency follow-up observations, or later observations with the same sensitivity. It has been suggested that some RRATs would be detected in standard FFT approach if they are closer and had higher signal-to-noise (S/N) observations \citep{wsrw06}. However, for many RRATs, it is necessary to measure times-of-arrival and determine timing solutions by using single pulses instead of the commonly used folded profiles.

Since their initial discovery, many theories have been put forward to explain why RRATs show different emission behavior from other pulsars. These include radio emission being disrupted by fallback of supernova material \citep{li06}, trapped plasma being released from radiation belts \citep{lm07}, and circumstellar material affecting the charge density in the magnetosphere \citep{cs08}. Alternatively, RRATs may be just one part of the neutron star intermittency spectrum, which sits as the extension of nulling pulsars with extremely high nulling fractions \citep{bur13}. In order to better understand their relation to other pulsars and the nature of the emission, we require the discovery of additional RRATs and, most importantly,  long-term monitoring and timing observations.

In this paper, we introduce our observations and data analysis methods for eight RRATs, followed by their timing solutions and the results from other studies probing other emission parameters. We also conduct a study of the RRAT population based on these and other timing solutions in order to find the similarities and differences in the spin-down properties of RRATs and normal pulsars.

\section{Discovery and Observations}

\begin{table}[!h]
\vspace{-1.0cm}
\end{table}

\begin{table*}[!htbp]
\vspace{0cm}
\centering
\resizebox{\textwidth}{!}{%
\hspace{-3cm}
\begin{tabular}{lccccccccccc}
\tableline\tableline
PSR Name  & Telescope & Data Machine & Frequency & Bandwidth & Sample Time & Time Span & Number  & Number  & Burst Rate & Mean Flux Density\\
      & & & (MHz) & (MHz) & ($\mu$$s$) & (Years) & of Observations & of TOAs & ($\rm{hr^{-1}}$) & for Single Pulses (mJy)\\
\tableline
J0735$-$6302  & Parkes & BPSR & 1400 & 256 & 64 & 2.06  & 22  & 304  &  39.65      & 0.4\\
J1048$-$5838  & Parkes & SCAMP & 1400 & 256 & 100 & 15.2  & 52  & 207  &  3.98 	& 2.4\\
J1226$-$3223  & Parkes & BPSR & 1400 & 256 & 64 & 1.94  & 19  & 360  &  41.26 	& 2.8\\
J1623$-$0841  & GBT & GUPPI & 350/820 & 100/200 & 245.76 & 2.43  & 24  & 1202  &  35.77 	& 11 (820 MHz); 84 (350 MHz)\\
J1739$-$2521  & GBT & GUPPI & 820 & 200 & 491.52 & 1.60 & 25  & 321  &  22.64 	& 49\\
J1754$-$3014  & GBT & GUPPI & 350/820 & 100/200 & 245.76 & 1.56  & 25  & 550  &  79.79	& 21 (820 MHz); 410 (350 MHz)\\
J1839$-$0141  & GBT & GUPPI & 820 & 200 & 491.52 & 2.50  & 38  & 386  &  15.61 	& 18\\
J1848$-$1243  & GBT & GUPPI & 820 & 200 & 245.76 & 1.80  & 29  & 393  &  30.35 	& 13\\
\tableline
\end{tabular}}
\caption{Observing parameters for eight RRATs; Here we list information of observations for the eight RRATs. The sample times used in GBT data for PSRs J1623$-$0841, J1739$-$2521, J1754$-$3014, J1839$-$0141, and J1848$-$1243 are converted from original according to the predicted smearing and scattering in order to reduce the size of the raw data. The mean peak flux densities are the average value of peak flux density.\label{tab1}}
\end{table*}

Two of the RRATs discussed in this paper, PSRs J0735$-$6302 and J1226$-$3223, were discovered in re-analyses of the 2009 southern-sky high Galactic latitude survey data \citep{bb09,jbo+09}, five of them (PSRs J1048$-$5838, J1754$-$3014, J1839$-$0141, and J1848$-$1243) were discovered in re-analyses of the 2001 Parkes Multibeam Pulsar Survey data \citep{man01a,kel+09}, and PSR J1623$-$0841 was discovered through the 2007 GBT 350 MHz drift-scan survey \citep{blr+13}. Furthermore, here we report the RRAT PSR J1739$-$2521 which was also discovered from re-analyses of the Parkes Multibeam Pulsar Survay data but not being reported before. We have carried out long-term timing and monitoring observation utilizing the Green Bank Telescope (GBT) with the GUPPI backend \citep{drd+08} and the Parkes Telescope with BPSR (and SCAMP) backend \citep{kjs+10}. Information on these observations is listed in Table \ref{tab1}.

\section{Analyses and Results}

We performed several analyses varying from preparing the raw observation data to the analysis of pulse properties, to measuring phase-connected timing parameters. Here we describe those steps in detail.

\subsection{Single-pulse Search}

\begin{figure}[!tbh]
\vspace{0.2cm}
\begin{center}
\includegraphics[trim=220 20 200 70,width=6.8cm,angle=0]{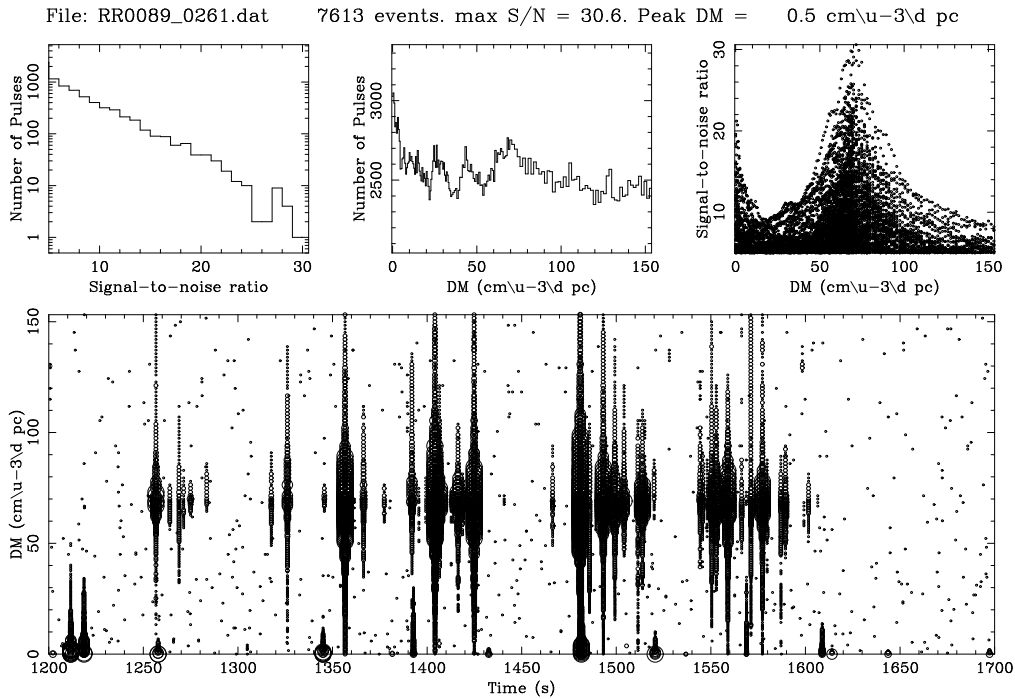}%
\vspace{-0.4cm}
  	\caption{Single-pulse search plot for PSR J1048$-$5838. The upper panels show numbers of pulses versus S/N (left) and DM (middle), and the pulse S/N distribution versus DM (right). The lower panel shows the search results versus both DM and time of detection; pulses are plotted with the radii of circles proportional to the S/N of pulses. This plot shows an 8-minute portion of a 60-minute Parkes 1.4 GHz observation during which the RRAT is ``on''. There are no pulses detected in its `off' phase. }
	\label{fig1}
  	\end{center}
\end{figure}

The first step in our timing analysis at each observation epoch is to identify which pulses are from the RRAT. Due to their sporadic nature, for many RRATs, we cannot use classical search algorithms based on Fourier techniques or folding. Therefore, we use the single-pulse search method to search for individual pulses with S/N above some threshold (in the case of our analysis, 5$\sigma$) in a number of trial-DM time series.  Here DM (dispersion measure) is the integrated column density of electrons along the line of sight. We do this by searching for pulses that are brighter at the DM of the RRAT than at zero DM using the `seek' command of the SIGPROC package.\footnote{See {\url{http://sigproc.sourceforge.net}}} Figure \ref{fig1} shows an example of the single-pulse search output for a portion of a nearly one-hour observation of PSR J1048$-$5838. We see very bright bursts that can be labelled as `real pulses' peaking at the source DM of 69 pc cm$^{-3}$, indicating they are astrophysical. The majority of signals that are due to local radio frequency interference (RFI) peak at DM of 0 pc cm$^{-3}$. The pulsar only turns `on' for six minutes of this observation. The total numbers of detected pulses and rates of pulse detection for all our target RRATs are listed in Table \ref{tab1}. For follow-up timing observations, we simplify this process by using only two trial-DM time series at the target RRAT DM and at zero DM, and check the pulses discovered to ensure that they are in phase with the pulse period.

\subsection{Spin Period and Time-of-arrival Calculation}
 
\begin{figure}[!h]

\begin{center}
\includegraphics[trim=50 360 0 80, width=11.5cm]{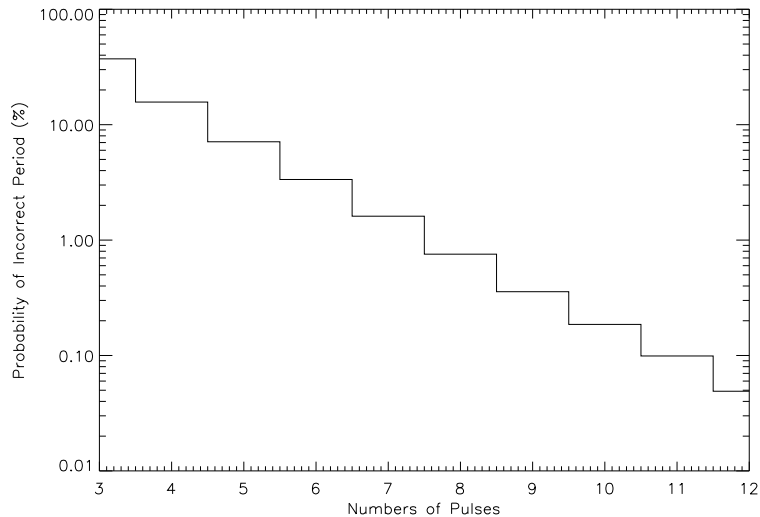}%
\vspace{-0.2cm}
  	\caption{Probability of calculating an incorrect period from the greatest common denominator method vs. number of pulses detected (in logarithmic scale). This simulation is for a 8-minute observation with 100 $\mu$s sample time. However, further testing shows that neither spin period, sample time nor observation time significantly affect this probability. The probability of calculating an incorrect period from three pulses in a 12 hr observation is 37.40\%, comparing with 37.01\% in 30 minutes.}
 	 \label{fig2}
  	\end{center}
\end{figure}

Prior to getting a timing solution, we must first calculate the spin period. We do this by measuring the differences between pulse arrival times and calculating the greatest common denominator of these differences. With a small number of detected pulses in an observation, there is a probability that this will be an integer multiple of the actual spin period. In order to find the probability of measuring the true period given some number of randomly distributed pulses, we created a large number of simulated RRAT-like timeseries with given sample time, period and pulse number, and calculate the greatest common denominator of the differences. The result shows that the number of pulses largely determines the probability of calculating an incorrect period (which is an integer multiple of the true period). The relationship between this probability and the number of pulses detected is shown in Figure \ref{fig2} \citep[also see][]{mll+06}. If eight or more pulses are detected, the probability of this method determining the correct period is greater than 99\%. Note that this calculation assumes that all of the pulses used for the calculation are actually from the source; if RFI pulses are mistakenly included in the calculation, the period will most likely be incorrect regardless of the number of pulses. Fortunately, the accuracy of the period can be later confirmed through the timing process.

Once a period is known, we bin the data into single pulses with 256 or 512 bins (depending on the pulse width) to generate profiles for these single pulses, and double check whether the detected pulses are real through visual inspection. If the period is known, the single-pulse profiles should all peak at roughly the same spin phase (i.e. within the span of the full profile, which varies from source to source). For the pulsars that we are only able to time through single pulses, we measure times-of-arrival (TOAs) as the arrival times of each single-pulse peak instead of through cross-correlation with a pulse template because the shapes of individual pulses can vary dramatically. For some RRATs that are more `pulsar-like' and less sporadic, we measure integrated TOAs by folding the `on'-phase data (typically on timescales of minutes) or even folding all the data for each observation, and use a template profile based on a high S/N observation for cross-correlation.

\subsection{Dispersion Measure Fitting}

A broadband signal will exhibit a frequency-dependent time delay caused by dispersion due to traveling through the interstellar medium (ISM). The time delay $t_{DM} \propto$ DM$/\nu^2$, where $\nu$ is the frequency of observation. The DM can be used to estimate the distance to a pulsar \citep{cl02}. To calculate a precise DM, we measure TOAs in multiple frequency bands and fit for the delay using TEMPO2 \citep{hem06}. 

We were able to fit DM for three RRATs. For PSRs J1623$-$0841 and J1754$-$3014, we fit TOAs at two frequencies, 350 MHz and 820 MHz. For PSR J1048$-$5838, we only have observations centered at 1.4 GHz with a 256 MHz bandwidth. To perform the DM fitting for this pulsar, we dedispersed the raw observation data into four sub-bands, so that each band has a bandwidth of 64 MHz. Then we implemented the single-pulse search and DM fitting. For the other RRATs, the single pulses have too low S/N when the data are split into sub-bands to provide reliable DM measurements.

\section{Timing}

\begin{figure}[!htb]
\hspace*{0.4cm}
\vspace*{-0.0cm}
\includegraphics[width=8.4cm]{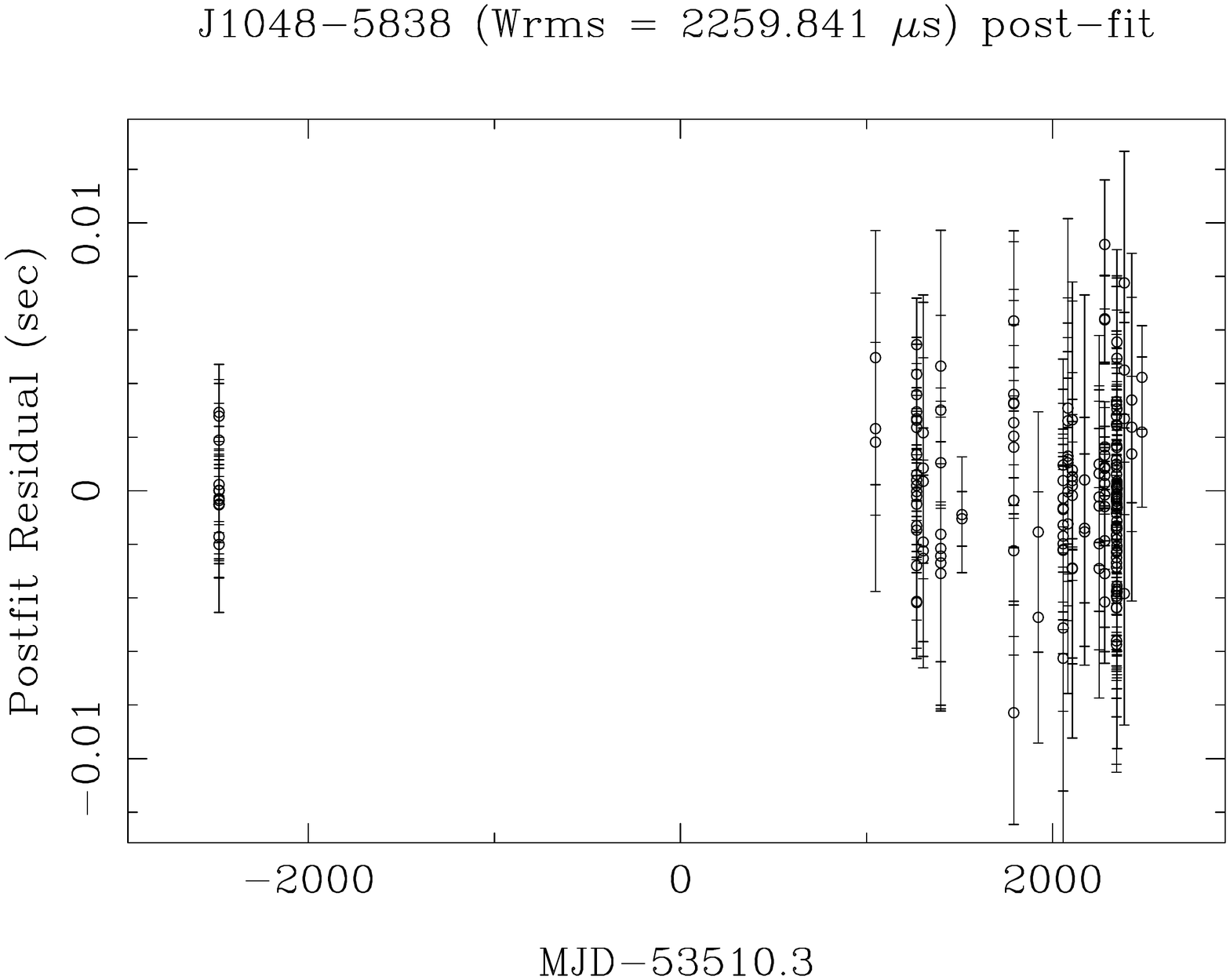}
\hspace*{-0.4cm}
\includegraphics[width=8.4cm]{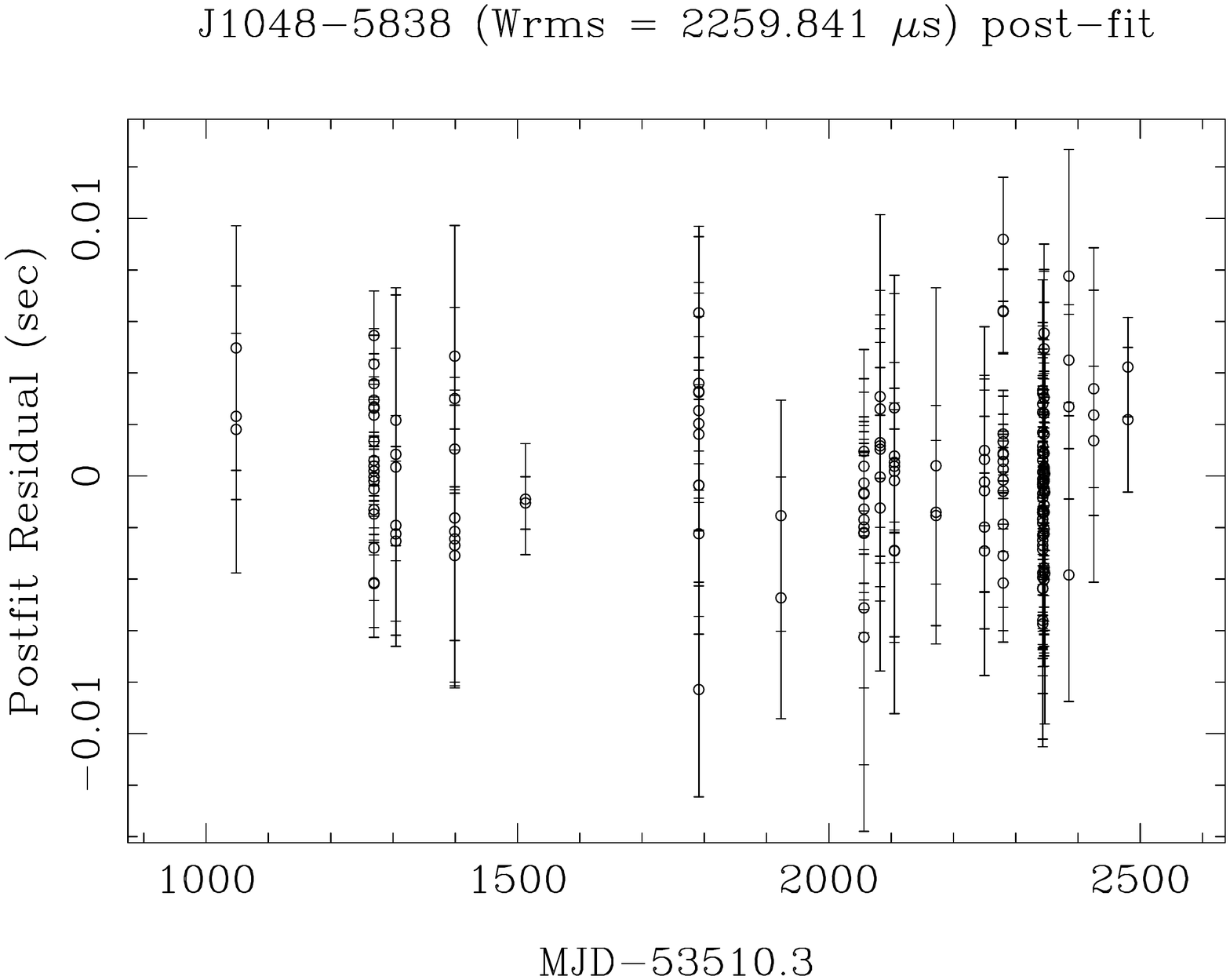}\\
\vspace{-0.7cm}
\caption{Left: all timing residuals from the four-year observation of PSR J1048$-$5838 starting in 09/2008 (see Table \ref{tab1}). Right: residuals excluding the discovery in 1998. Some data points vary by more than the error bar; this is likely due to pulse jitter.\label{fig3}}
\end{figure}

We calculate a timing model and fit for timing residuals using the pulsar timing software TEMPO2 \citep{hem06}. The full solutions are listed in Table \ref{tab2}, and some remarks about the timing parameters of eight RRATs are provided here. Most of the RRATs we timed have been observed for the time span of one to three years, but PSR J1048$-$5838 has an exceptionally long span of observation. This includes four years of post-discovery timing observations and a total 15-year span including the discovery, with our solution producing phase connection very well (see Figure \ref{fig3}). The root-mean-square (RMS) values of post-fit residuals for all our RRATs range from 0.8 to 7.6 ms (see Table \ref{tab2}).  

\begin{table}[!ht]
\vspace{-0.8cm}
\end{table}

\begin{table*}[!htbp]
\vspace{0cm}
\centering
\resizebox{\textwidth}{!}{%
\hspace{-3cm}
\begin{tabular}{lccccccccccc}
\tableline\tableline
PSR Name & R.A. & Decl. & $P$ & $\dot{P}$ & RMS & DM & Epoch & Width & $B$ & $\rm{\dot{E}}$ & $\tau$\\
 & (J2000) & (J2000) & (s) & ($10^{-15}$ s s$^{-1}$) & (ms) & (pc cm$^{-3}$) & (MJD) & (ms) & ($10^{12}$ G) & ($10^{31}$erg s$^{-1}$) & (Myr)\\
\tableline
J0735$-$6302 & 07:35:05(20)	& $-$63:02:00(40) 	& 4.862873966(3)	& 159(5) 	& 6.38 	& 19.4 		& 56212 & 24	& 28 	& 5.5 	& 0.5\\
J1048$-$5838 & 10:48:12.57(2)	& $-$58:38:18.58(15) 	& 1.231304776631(3) 	& 12.19369(7) 	& 2.22 	& 70.7(9) 	& 53510	& 7   & 3.9 	& 26 	& 1.6\\
J1226$-$3223 & 12:26:45.9(4)	& $-$32:23:14(5) 	& 6.1930029826(6) 	& 7.68(11)   	& 7.55 	& 36.7 		& 56114	& 57	& 6.0 	& 0.08 	& 20\\
J1623$-$0841 & 16:23:42.711(10) & $-$08:41:36.4(5) 	& 0.503014992514(6) 	& 1.9582(6)	& 0.76 	& 60.433(16) 	& 55079	& 13	& 1.0 	& 61 	& 4.1\\
J1739$-$2521 & 17:39:32.83(10)	& $-$25:21:02(20) 	& 1.8184611641(2) 	& 0.29(3)	& 4.85 	& 186.4 	& 55631	& 68	& 0.7 	& 0.7 	& 99\\
J1754$-$3014 & 17:54:30.08(5) 	& $-$30:14:42(6) 	& 1.3204902915(3) 	& 4.424(12)	& 3.61 	& 99.38(10) 	& 55292	& 62	& 2.4 	& 7.6 	& 4.7\\
J1839$-$0141 & 18:39:07.03(3) 	& $-$01:41:56.0(9) 	& 0.93326564072(6) 	& 5.943(3)	& 1.80 	& 293.2(6) 	& 55467	& 17	& 2.4 	& 29 	& 2.5\\
J1848$-$1243 & 18:48:17.980(8) 	& $-$12:43:26.6(5) 	& 0.41438334869(2) 	& 0.440(2)	& 1.36	& 88.0		& 55595	& 9	& 0.4 	& 24	& 15\\
\tableline
\end{tabular}
}
\caption{Timing solutions and derived parameters for all eight RRATs: Right Ascension, Declination, spin period, period derivative, root-mean-square of residuals, DM, epoch of period measurement, width of composite profile, magnetic field, spin-down energy loss rate and characteristic age are listed. Here the width is calculated at 50\% of the peak intensity (W50), the magnetic field is at the pulsar surface and assumes alignment between spin and magnetic axis (\(B = 3.2 \times 10^{19} \sqrt{P\dot{P}}\)). The pulsar spin-down luminosity is calculated by \(\dot{E} = -4 \times 10^{46} \dot{P}/P^3\).  \label{tab2}}
\end{table*}

\vspace{-0.3cm}

\subsection{Position}

Most of the RRATs have timing-derived positions within the original discovery beam, such as the 3$^{\prime}$ difference for PSR J1048$-$5838's position from the center of Parkes's 1.4-GHz beam, which is roughly 14$^{\prime}$ in size. However, for PSR J1623$-$0841, the final TEMPO2 fitted position was 20$^{\prime}$ away from the discovery position (outside the GBT's 820 MHz beam) as an offset in position during an observation acts like a decrease in gain of the telescope, thus lowering the sensitivity. A timing solution was only attainable with the increased sensitivity of GUPPI (with twice the bandwidth of the original SPIGOT backend used for the discovery observation) and a dense set of observations to obtain a phase coherent timing solution at 350 MHz.

\subsection{Period}

When PSR J1739$-$2521 was discovered, it was inferred to have a period of $\sim$1.2 s. Further analysis indicates that the true period is 1.82 s. The reason for the initial incorrect period is because the profile has two peaks and both peaks produce single pulses that are detected in the single-pulse search. This causes the algorithm that determines the period from single pulses to fail. Each detection must be looked at by eye to determine if a single pulse actually produced two detections in the single-pulse search.

PSR J1754$-$3014 was originally reported in \citet{mll+06} with a period of 0.42 s and a DM of 98(6) pc cm$^{-3}$. In \citet{kkl+11} it was reported to have a period of 1.32 s and a DM of 293(19) pc cm$^{-3}$. Here we report the same period as in \citeauthor{kkl+11} and a DM of 99.38(10) pc cm$^{-3}$. The difference between our period and \citet{mll+06} is due to the misidentification of a terrestrial radio pulse as an astrophysical pulse \citep{kkl+11}. The difference between our DM and \citet{kkl+11} is due to a formatting error in their paper, resulting in PSR J1754$-$3014's DM being confused with PSR J1839$-$0141's DM. 

PSR J1839$-$0141 was originally reported in \citet{mll+06} to have a period of 0.932 s and a DM of 307(10) pc cm$^{-3}$. Here it is reported with a period of 0.933 s and a DM of 293.2(6) pc cm$^{-3}$. The difference in DM may be due to the coarse frequency resolution of the PMPS and is less than a 2$\sigma$ difference from the discovery DM. The difference between the discovery period and the period reported here is much larger than what would be produced by the measured $\dot{P}$. It was only seen in one of 10 observations reported in \citet{mll+06} and has never been detected in 38 observations with the Parkes telescope since its discovery. It is possible that the low S/N of the discovery pulses is responsible for the significant differences between the calculated periods.

\subsection{Pulse Profiles}

Many RRATs cannot be detected by summing all the rotations over an observation. Therefore, to create integrated pulse profiles, the most straightforward way is adding all detectable single pulses after phase corrections from the timing model. This provides us with relatively stable pulse profiles. For some RRATs that are less ``transient-like'', we can create integrated pulse profiles by folding the `on'-phase data for each observation, or even folding the entire observation as for other pulsars. 

\begin{figure}[!pb]
\center
\vspace*{-0cm}
\hspace*{-1.6cm}
 \begin{minipage}[b]{0.245\textwidth}
\centering \includegraphics[trim=0 0 100 100, width=7.8cm]{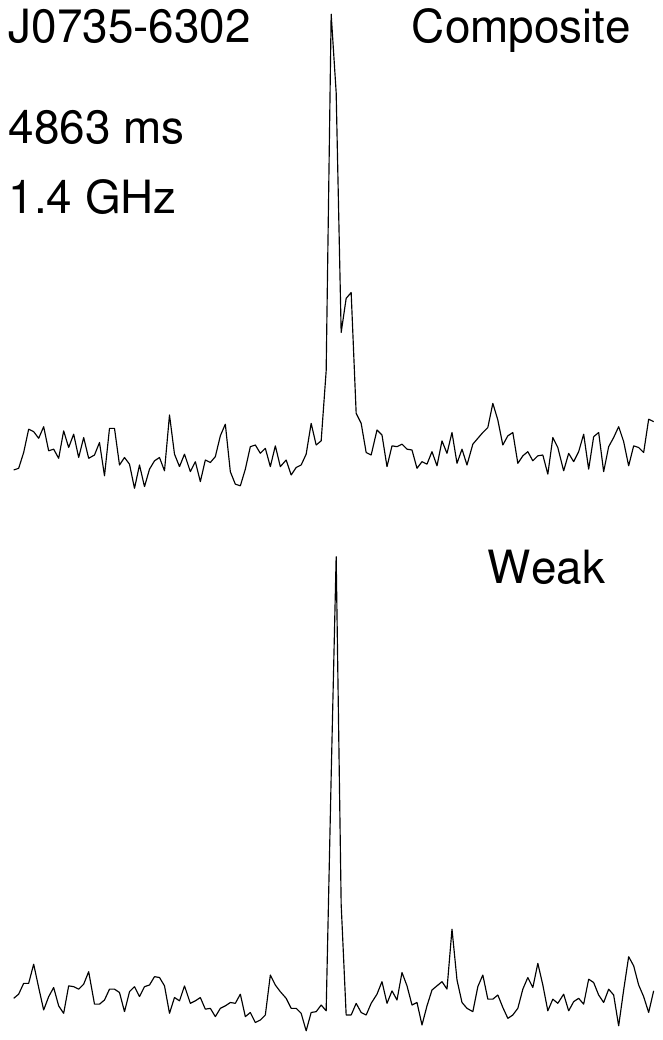}
 \end{minipage}
\hspace*{0cm}
 \begin{minipage}[b]{0.245\textwidth}
\centering \includegraphics[trim=0 0 100 100,width=7.8cm]{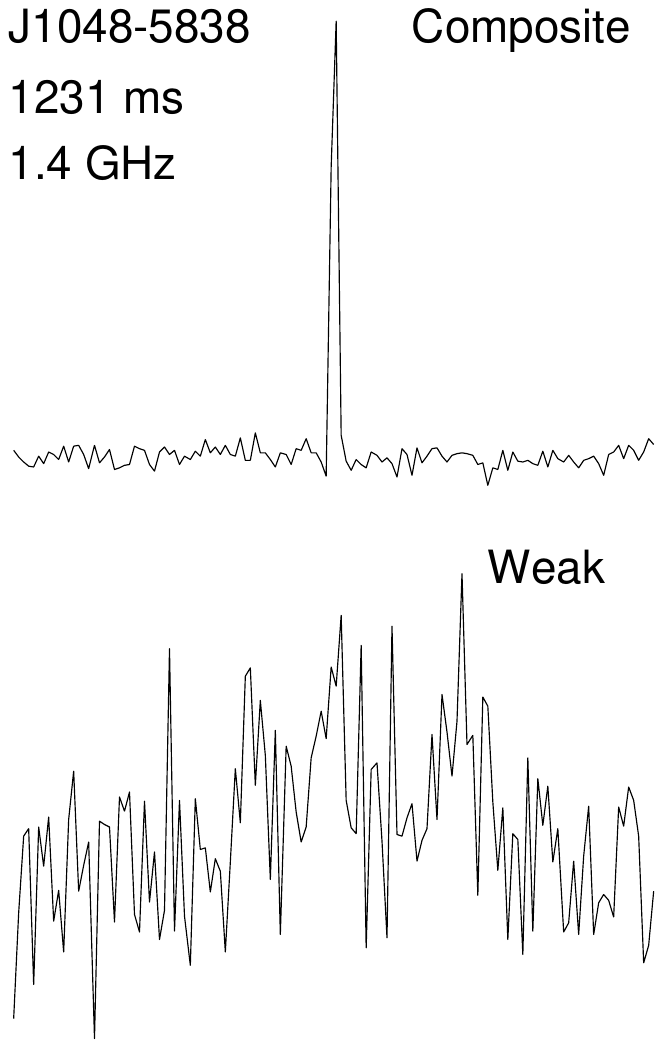}
 \end{minipage}
\hspace*{0cm}
 \begin{minipage}[b]{0.245\textwidth}
\centering \includegraphics[trim=0 0 100 100,width=7.8cm]{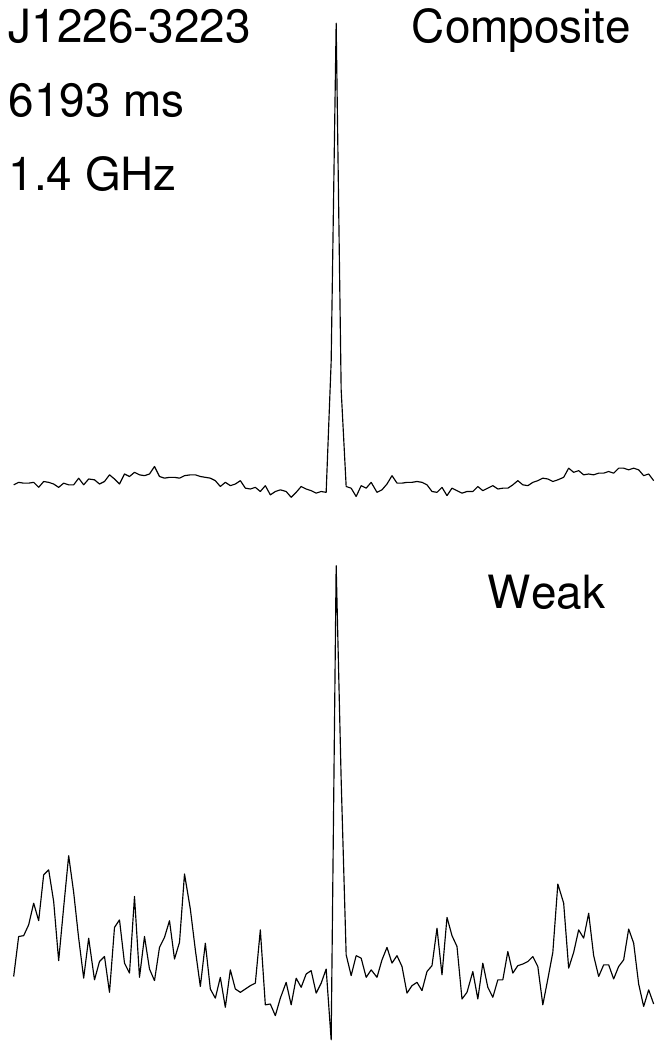}
 \end{minipage}
\hspace*{0cm}
 \begin{minipage}[b]{0.245\textwidth}
\centering \includegraphics[trim=0 0 100 100,width=7.8cm]{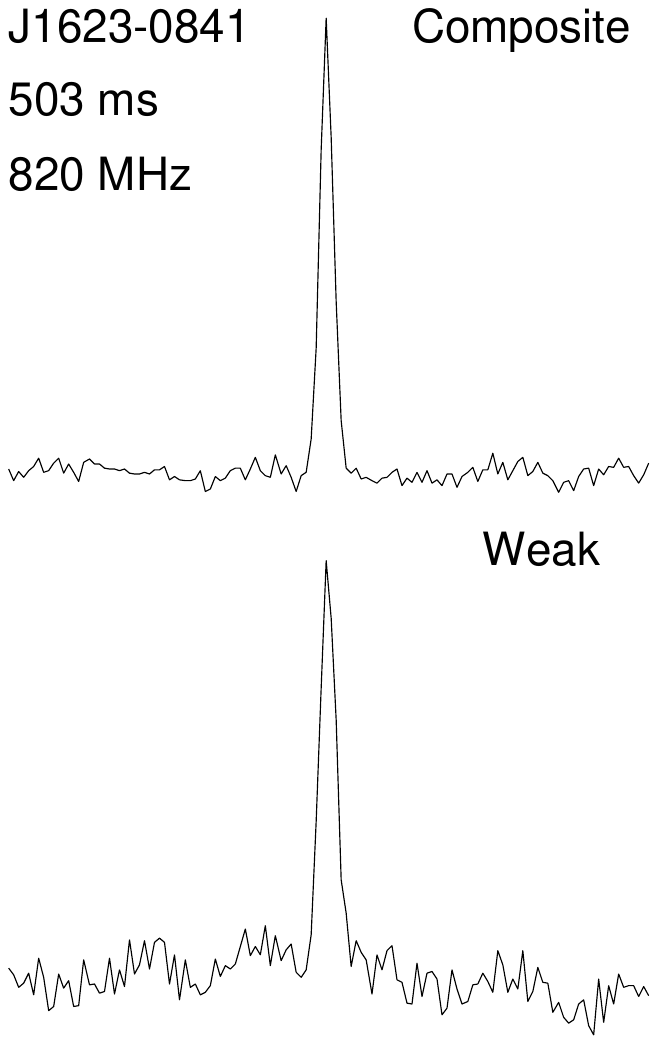}
 \end{minipage}
\hspace*{0cm}
\vfill
\vspace*{-5cm}
\hspace*{-1.6cm}
 \begin{minipage}[b]{0.245\textwidth}
\centering \includegraphics[trim=0 0 100 100,width=7.8cm]{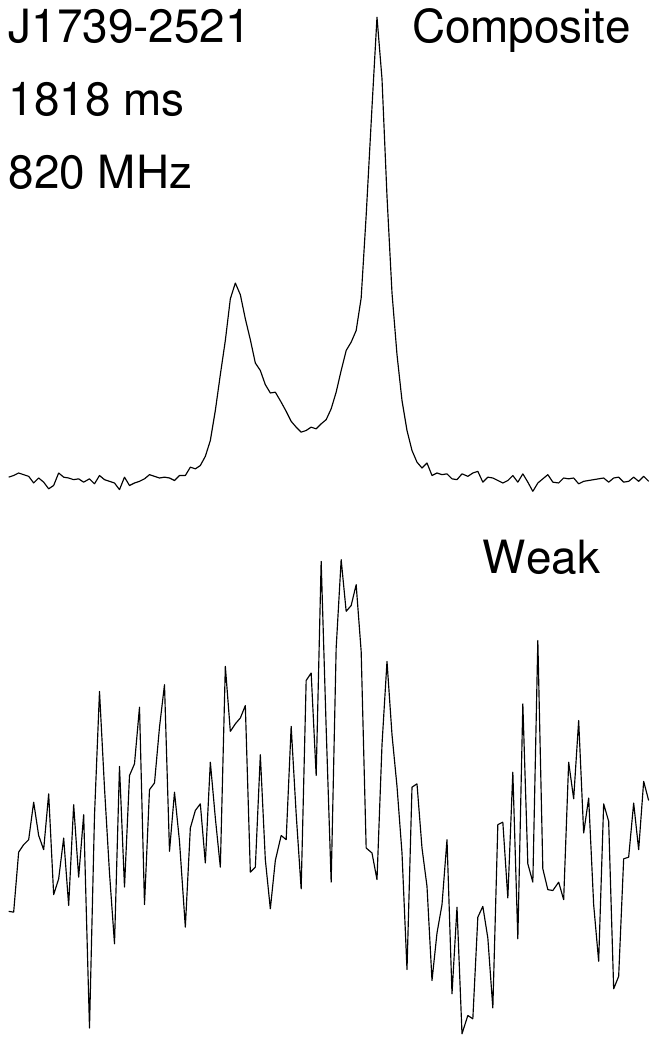}
 \end{minipage}
\hspace*{0cm}
 \begin{minipage}[b]{0.245\textwidth}
\centering \includegraphics[trim=0 0 100 100,width=7.8cm]{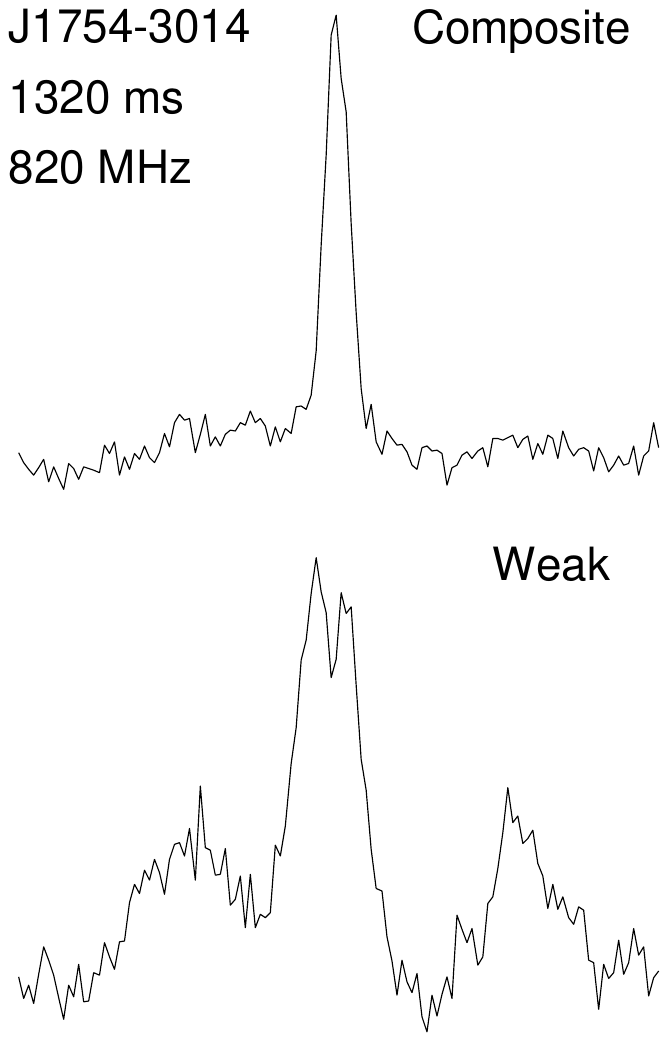}
 \end{minipage}
\hspace*{0cm}
 \begin{minipage}[b]{0.245\textwidth}
\centering \includegraphics[trim=0 0 100 100,width=7.8cm]{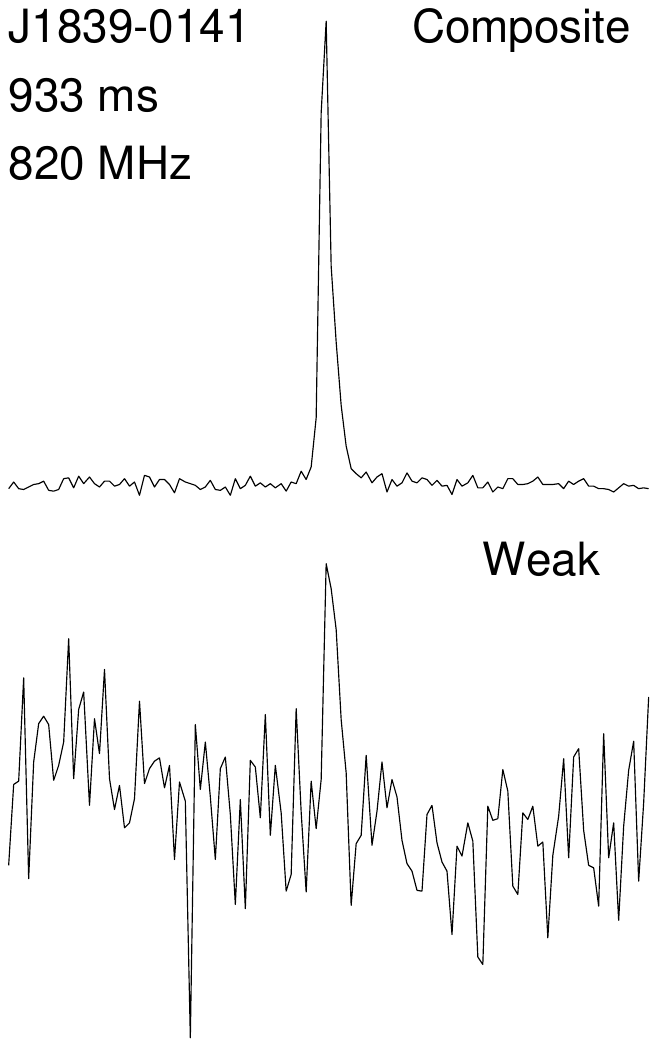}
 \end{minipage}
\hspace*{0cm}
 \begin{minipage}[b]{0.245\textwidth}
\centering \includegraphics[trim=0 0 100 100,width=7.8cm]{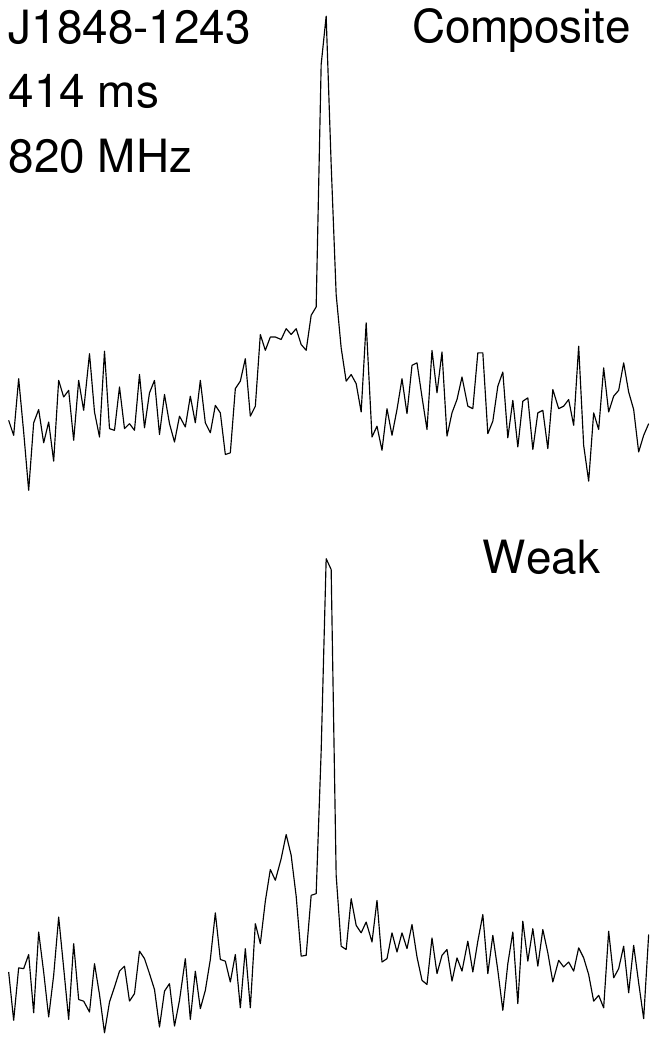}
 \end{minipage}
\hspace*{0cm}
\vspace*{-5.5cm}
\caption{Composite pulse profiles (upper) and weak pulse profiles (lower) of eight RRATs based on the observations described in Table \ref{tab1}. The spin periods and frequencies are listed. The profiles of PSRs J0735$-$6302, J1048$-$5838, and J1226$-$3223 are based on 1.4 GHz observations with the Parkes telescope, and are created by adding all detected individual single pulses. The others are from 820 MHz observations with the GBT, in which the pulse profiles of PSRs J1739$-$2521 and J1839$-$0141 are sums of data during the approximately minute-long time periods when the RRATs are ``on''. Those of PSRs J1623$-$0841, J1754$-$3014, and J1848$-$1243 are created by folding all of the data. The weak pulse profiles are generated by subtracting all detected single-pulse profiles from the composite profile created through folding all data in each observation, and indicate the fraction of weak pulses not being detected through the single-pulse algorithm. These profiles provide confirmation to the discussion of detectability correction (Figure \ref{fig_real}).} 
\label{fig4}
\end{figure}

Profiles of the eight RRATs are presented in Figure \ref{fig4} in the upper panels. Most of them are narrow (with duty cycles of less than 5\%), except for the double-peaked profiles of PSR J1739$-$2521 and possibly PSR J0735$-$6302 (shape not clear due to the low profile S/N). Here, the pulse profiles of PSRs J1623$-$0841, J1739$-$2521, and J1839$-$0141 are sums of data during the $\sim$minute-long time periods when the RRATs are `on'. The profiles of PSRs J0735$-$6302, J1048$-$5838, and J1226$-$3223 are sums of all detected individual single pulses. The remaining two RRATs (PSRs J1754$-$3014 and J1848$-$1243) were detected through single-pulse searches but are typically detectable in follow-up observations by folding all of the data, and therefore the integrated profiles consist of all data from all observations folded. The total integration time is 1478 s for J0735$-$6302, 255 s for J1048$-$5838, 2230 s for J1226$-$3223, 80 minutes for J1623$-$0841, 110 minutes for J1739$-$2521, 413.1 minutes for J1754$-$3014, 24.75 minutes for J1839$-$0141, and 835.5 min for J1848$-$1243, grant that the profile generation methods are different.

\vspace{0.2cm}
\section{Analysis}  

In this section, we explore the pulse amplitude distributions and weak pulse emission for the RRATs.

\subsection{Pulse Amplitude Distribution} \label{ssec:5.1}

One of the most important features of the single pulses is their amplitude distribution, since this is a direct probe of the internal emission mechanism of RRATs and provides a comparison with other pulsars. We take the peak S/N of each single pulse and calculate its flux density based on the radiometer noise calculation using the equation 
\begin{equation}\label{eqn1}
S_{\rm{sys}} = \frac{T_{\rm{sys}}}{G\sqrt{t_{\rm{obs}} n_{p}\Delta f}}
\end{equation} where $S_{\rm{sys}}$ is the system equivalent flux density, $T_{\rm{sys}}$ is the system noise temperature, $G$ is the gain of the telescopes (2 K/Jy for GBT and 0.6 K/Jy for Parkes), $t_{\rm{obs}}$ is the observation sample time, $\Delta f$ is the observed bandwidth, and $n_{p}$ is the number of polarizations (2 for all our observations). 

The peak S/N is calculated from the single-pulse profile
\begin{equation}\label{eqn2}
SNR_{\rm{peak}} = \frac{A_{\rm{peak}}}{\sigma_{\rm{off}}}
\end{equation} where $A_{\rm{peak}}$ is the peak amplitude of the profile, and $\sigma_{\rm{off}}$ is the standard deviation of the profile amplitude in the off-pulse region. 

\begin{figure}[!t]
\centering
\captionsetup{justification=centering}
\vspace*{-0.7cm}%
 \begin{minipage}[b]{0.45\textwidth}%
\centering \includegraphics[trim=100 250 10 0, width=6cm]{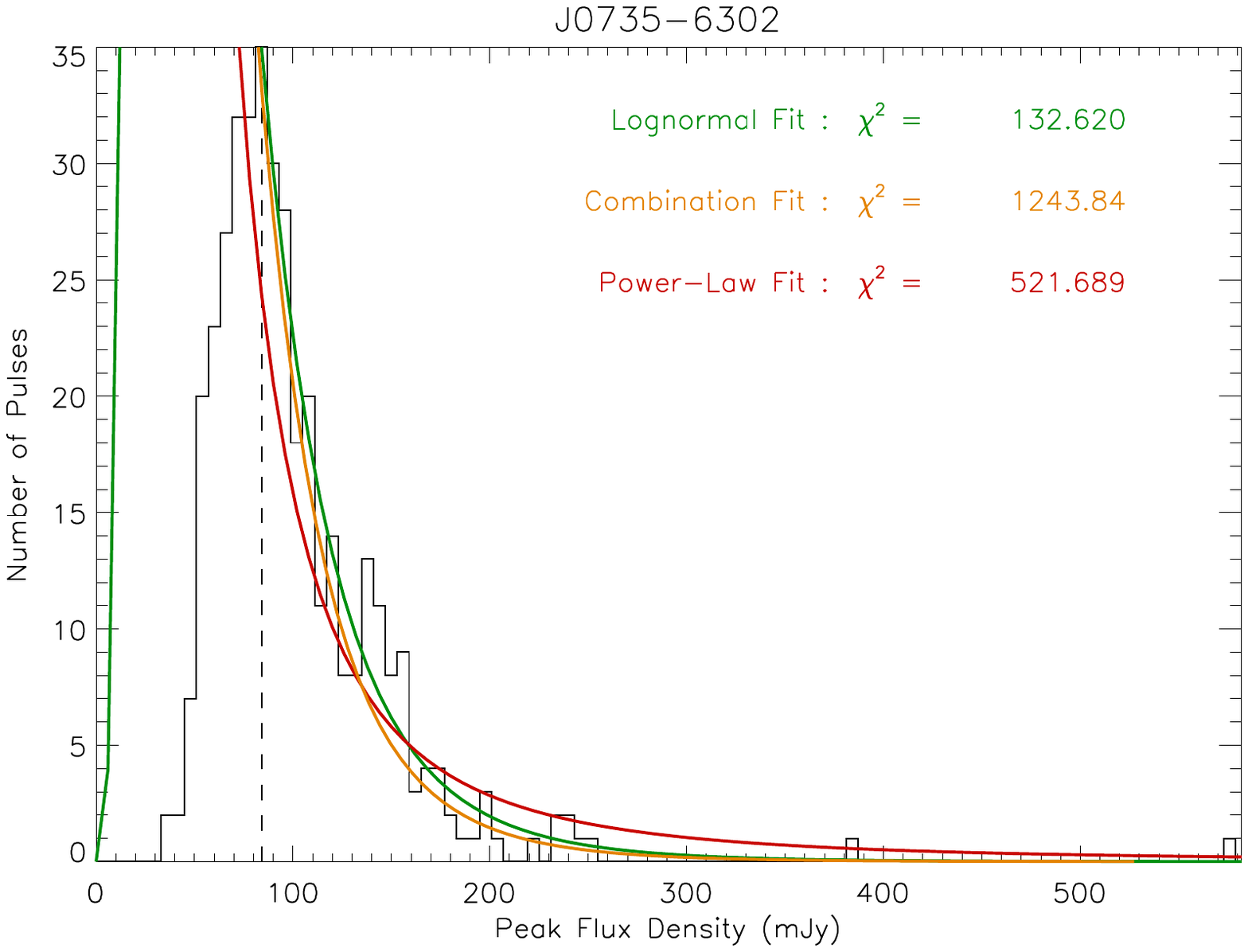}
\end{minipage}%
 \begin{minipage}[b]{0.45\textwidth}%
\centering \includegraphics[trim=100 250 10 0, width=6cm]{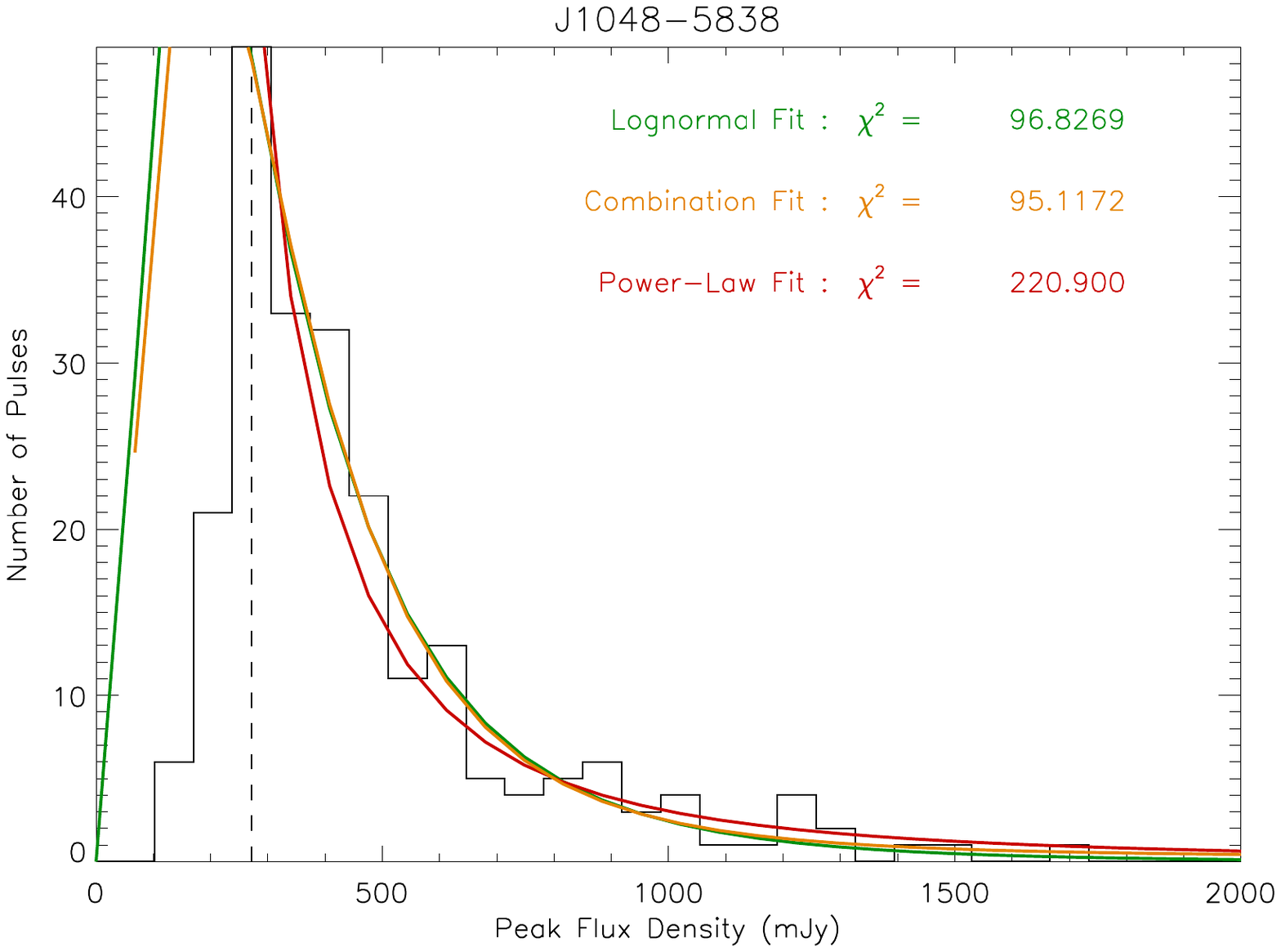}
 \end{minipage}
\vfill
\vspace*{-1.9cm}
 \begin{minipage}[b]{0.45\textwidth}%
\centering \includegraphics[trim=100 250 10 0, width=6cm]{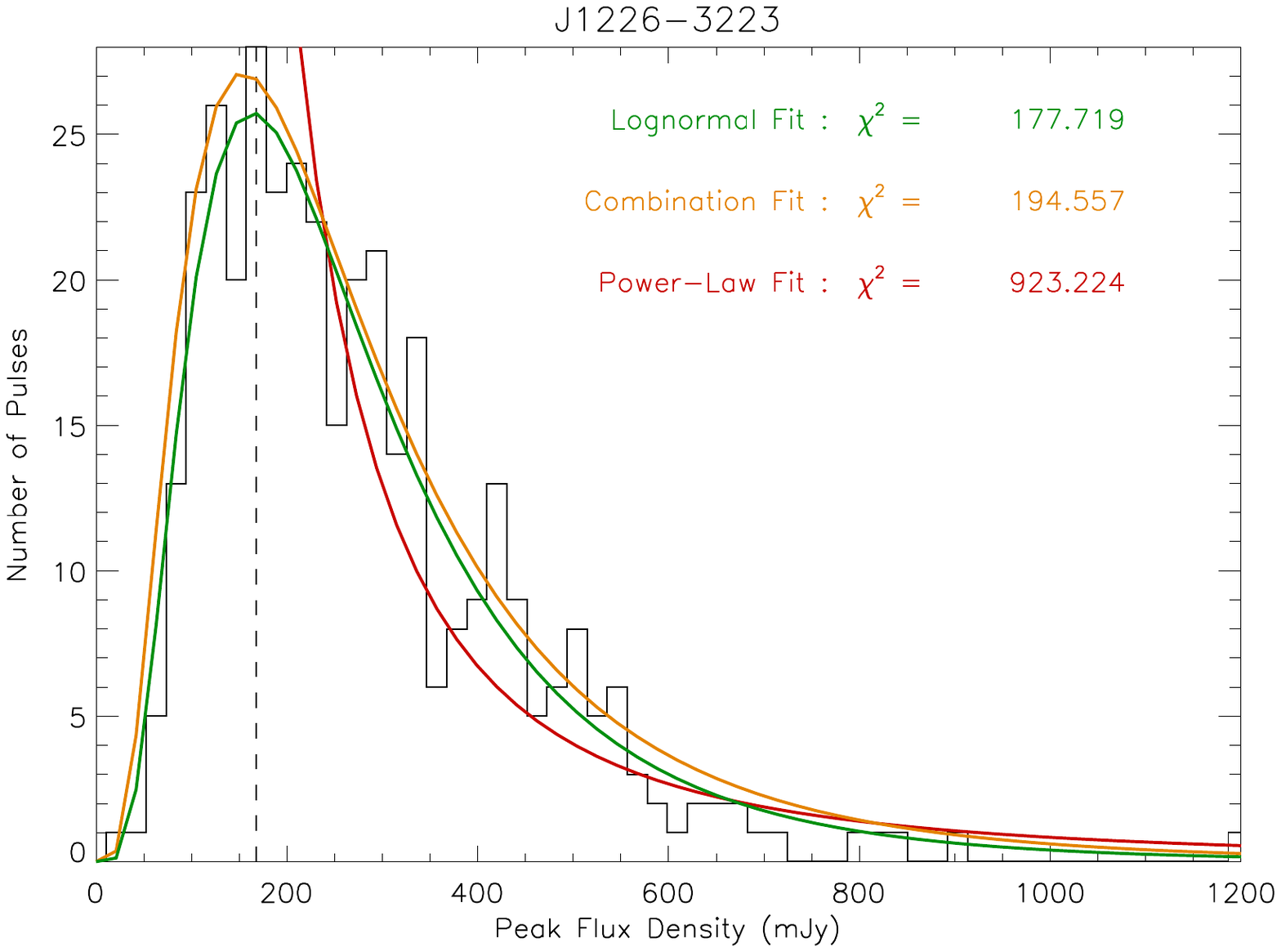}
 \end{minipage}%
 \begin{minipage}[b]{0.45\textwidth}%
\centering \includegraphics[trim=100 250 10 0, width=6cm]{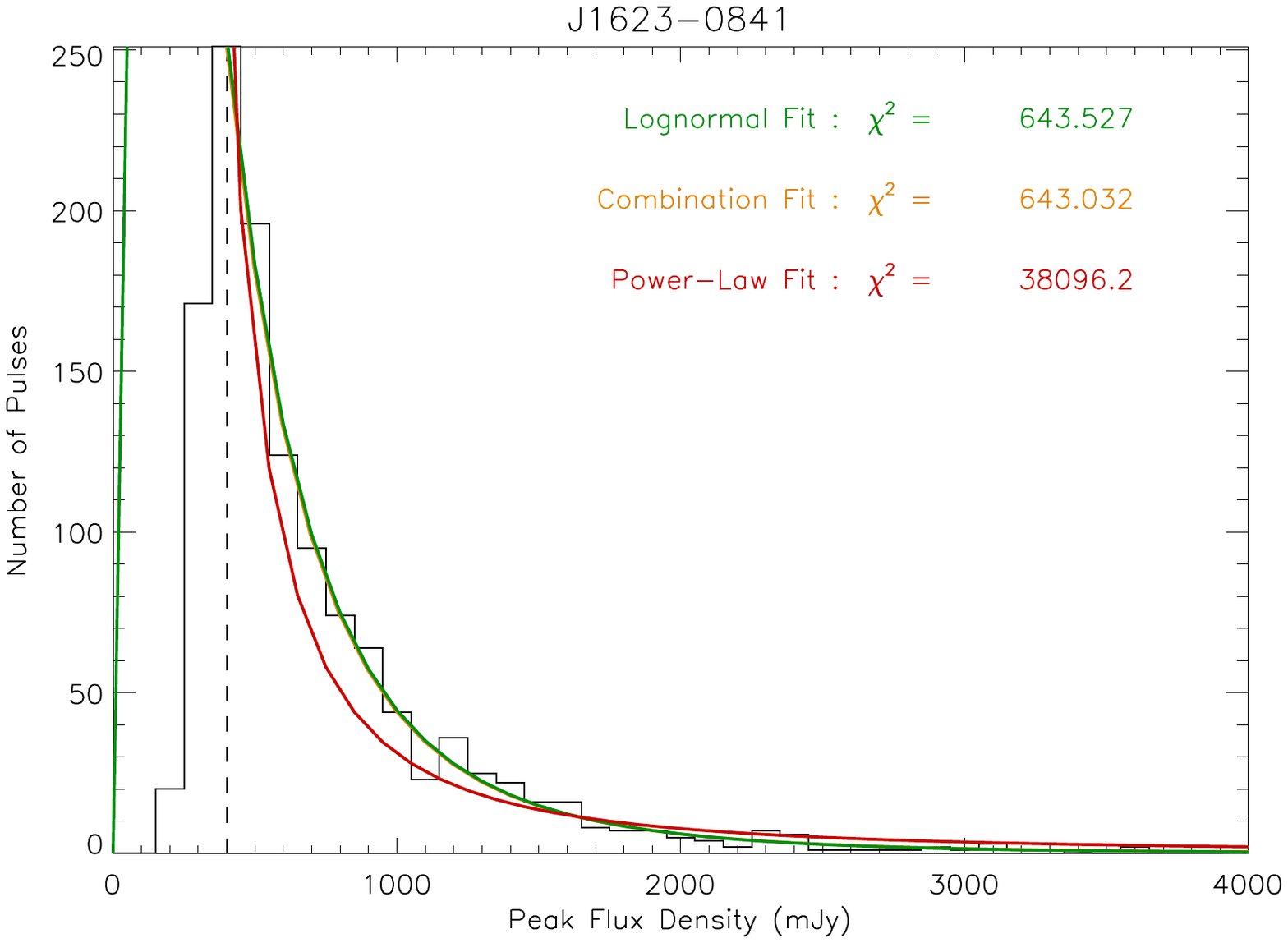}
 \end{minipage}
\vfill 
\vspace*{-1.9cm}
 \begin{minipage}[b]{0.45\textwidth}%
\centering \includegraphics[trim=100 250 10 0, width=6cm]{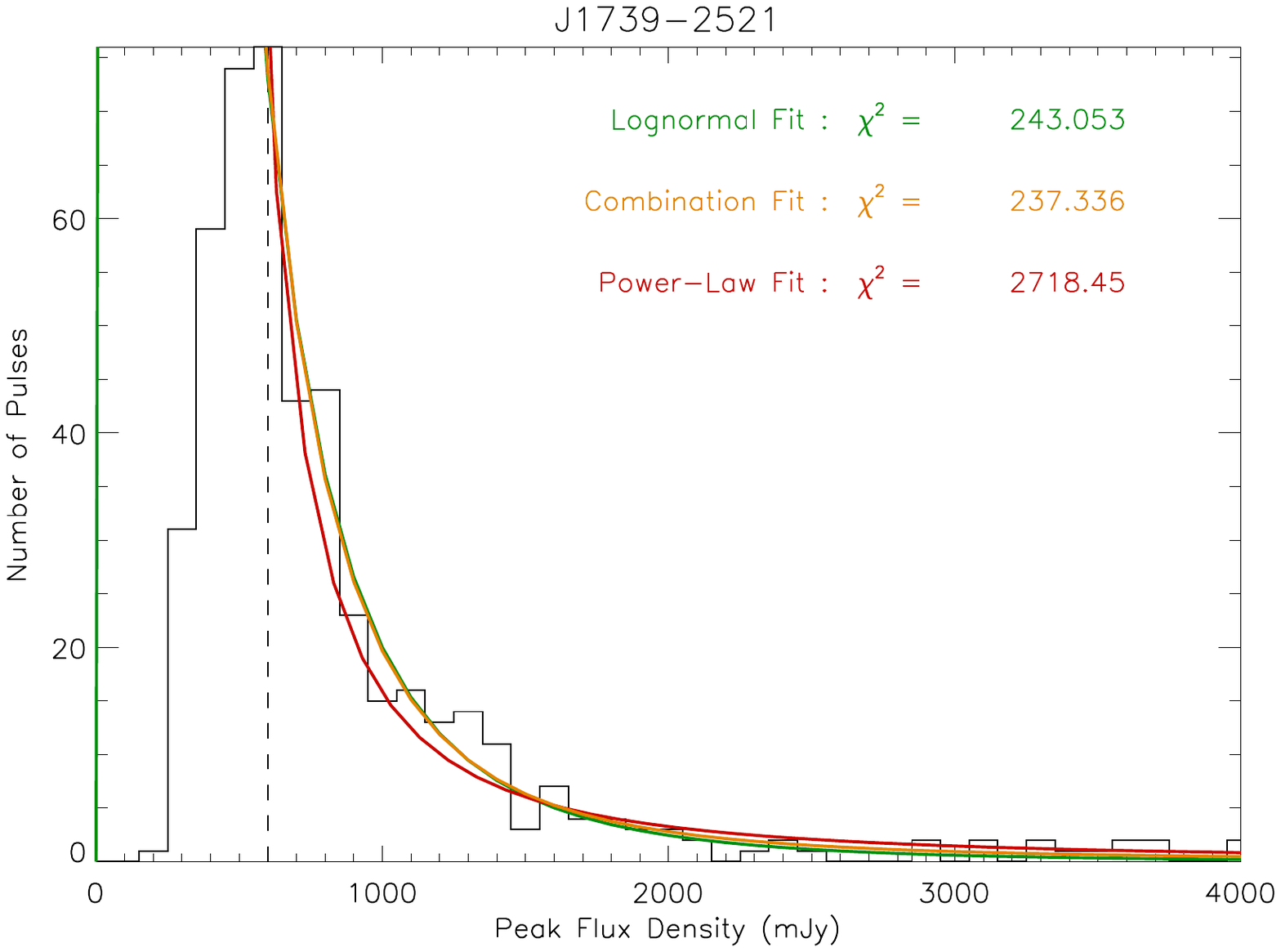}
 \end{minipage}%
 \begin{minipage}[b]{0.45\textwidth}%
\centering \includegraphics[trim=100 250 10 0, width=6cm]{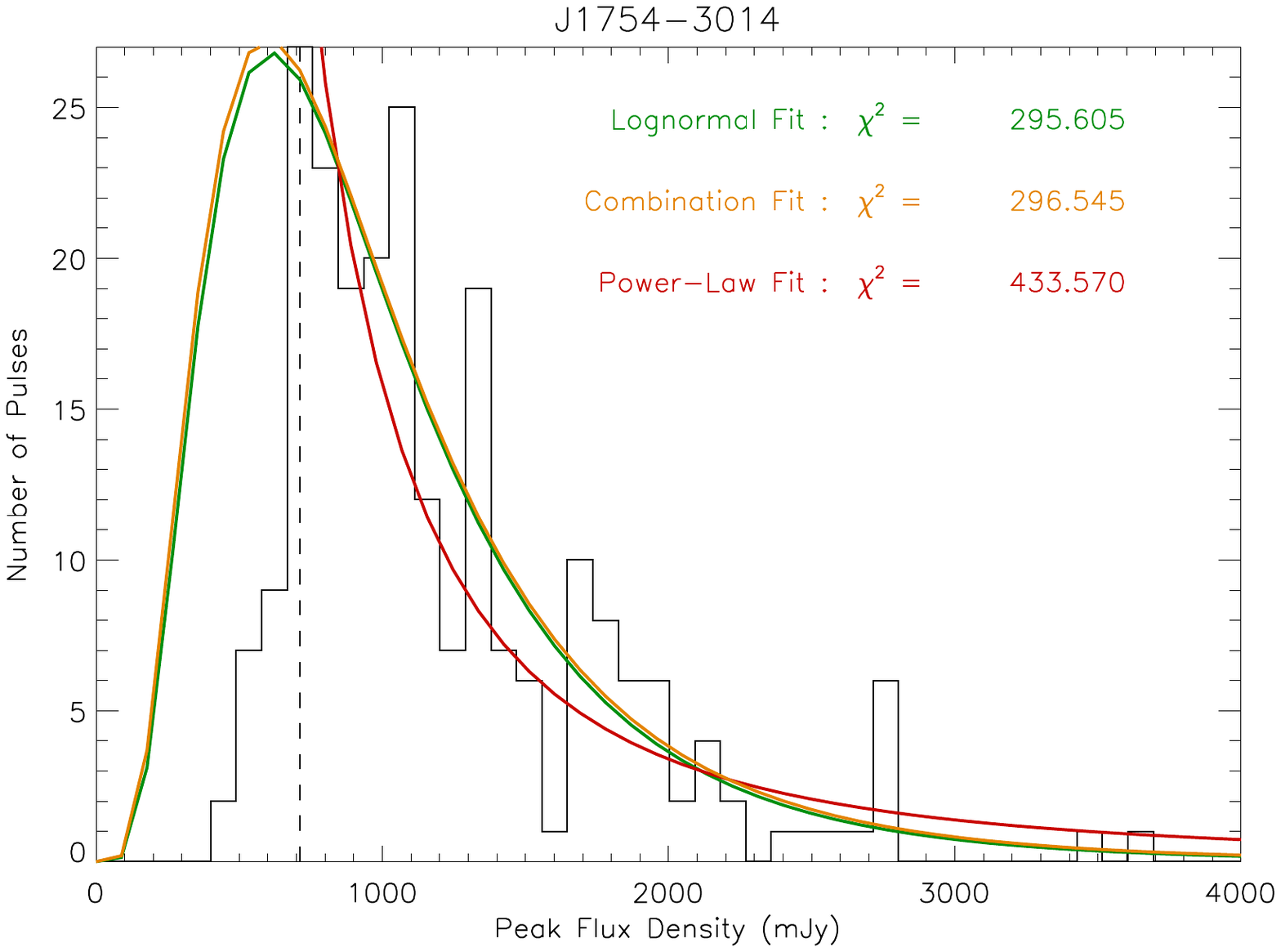}
 \end{minipage}%
\vfill
\vspace*{-1.9cm}
 \begin{minipage}[b]{0.45\textwidth}%
\centering \includegraphics[trim=100 250 10 0, width=6cm]{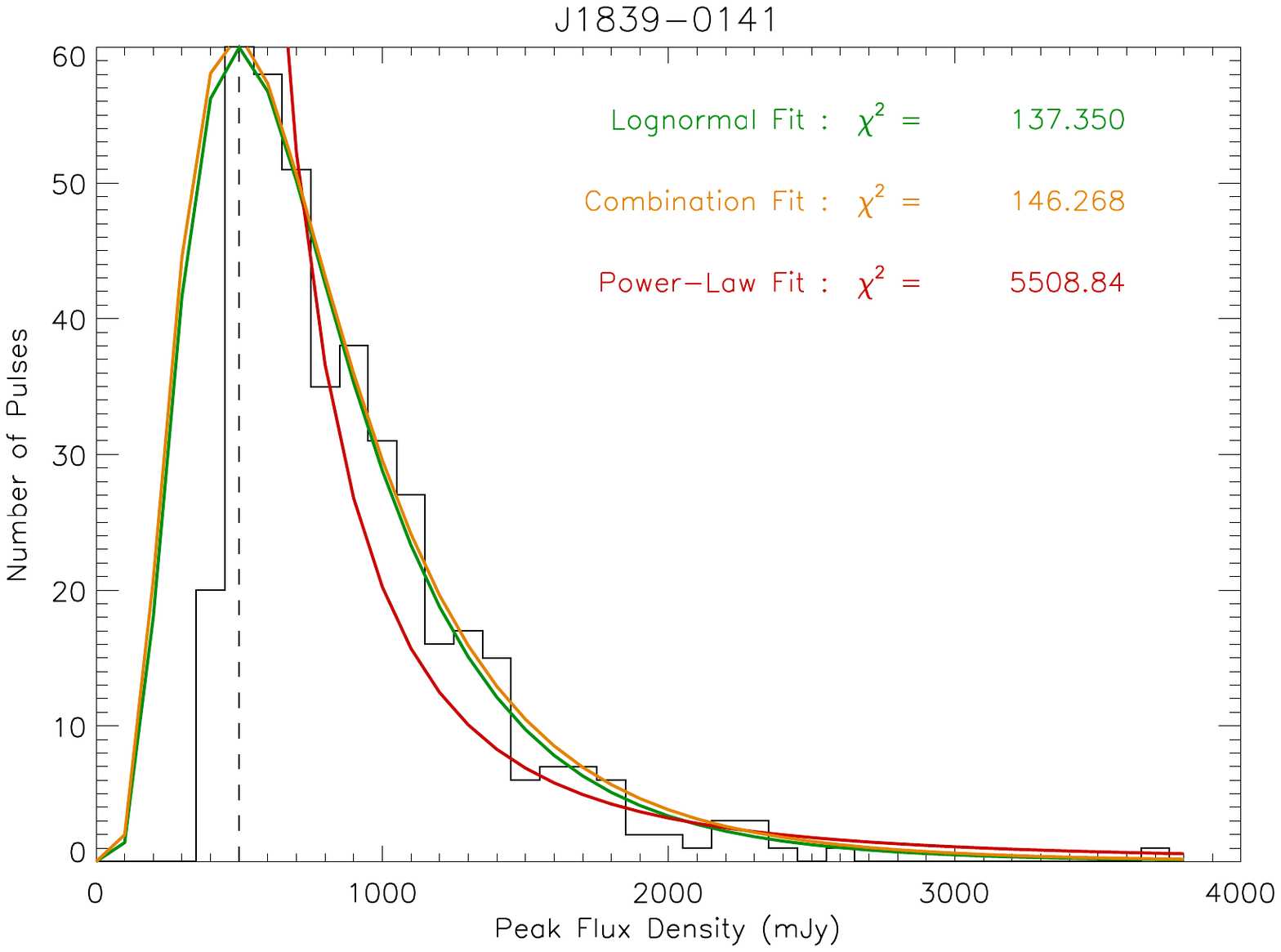}
 \end{minipage}%
 \begin{minipage}[b]{0.45\textwidth}%
\centering \includegraphics[trim=100 250 10 0, width=6cm]{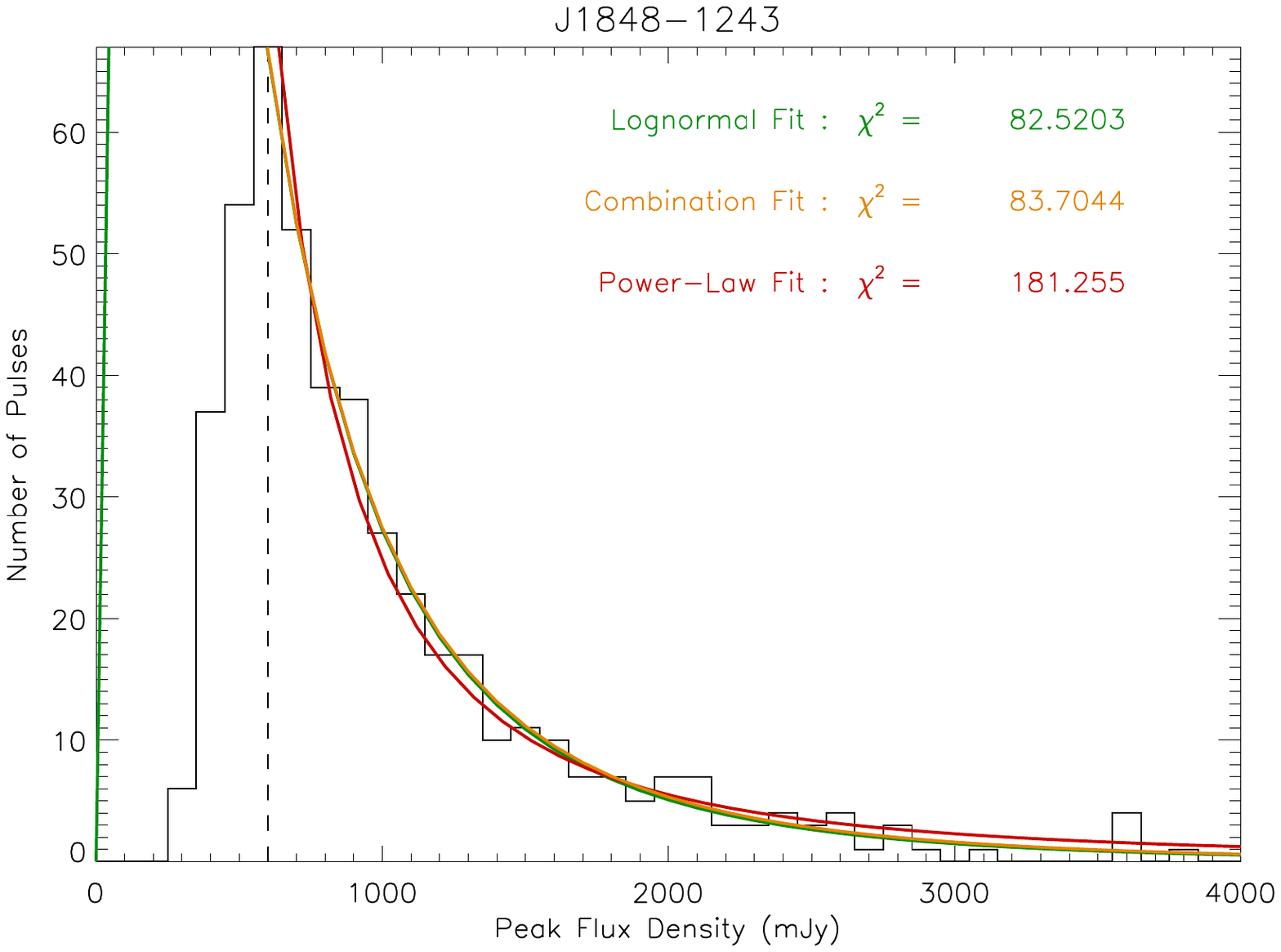}
 \end{minipage}%
\vspace*{-0.8cm}
\caption{Pulse amplitude distribution of eight RRATs. All peak flux densities are calculated based on the pulse S/N and radiometer noise. To reduce the influence from non-detected weak pulses, we fit only the distribution with flux densities above the distribution peak (dashed line) with three base functions (in different colors)  : log-normal distribution, power-law distribution and the combination of the two, with the $\chi^2$ results listed. From these results, we can see that the log-normal distribution fits all the eight RRATs very well and mostly dominates the combined function.}
\label{fig5}
\end{figure}

Then the peak pulse flux density can be calculated by
\begin{equation}\label{eqn3}
S_{\rm{pulse}} = SNR_{\rm{peak}} \times S_{\rm{sys}}
\end{equation}
The results are shown in Figure \ref{fig5}, where we fit these binned histograms to three different functions: log-normal distribution, power-law distribution, and the combination of the two. We choose these distributions because most pulsars show log-normal distributions and giant pulses show a power-law distribution \citep{mml+12}. The results indicate that a log-normal distribution overall provides the most accurate fit, but PSRs J1226$-$3223 and J1839$-$0141 show strong evidence for a power-law tail, similarly to the giant pulses emitted from some pulsars. Here the model of the intrinsic pulse energy distribution convolved with the noise distribution \citep{bjb+12} is not applied because of the relatively higher S/N of RRATs single pulses. Therefore, this would make only a very small correction and not affect the underlying derived distribution.

Note that fitting to the distributions above is only applied to the distribution for pulses with flux densities greater than some threshold due to the selection effects in detecting weak pulses. All of the RRATs show a turnover at low S/N in their amplitude distributions. In order to determine whether this turnover is intrinsic to the RRATs or due to decreased sensitivity to weak pulses, we need to check the detectability of pulses as a function of S/N. 

During the S/N calculation process, we were careful to convert between the different definitions of S/N used by the search algorithm and for creating our pulse amplitude plots (which we denote as ``search S/N" and ``profile S/N"). This difference is due to the different time resolutions of these two algorithms (the width of the boxcar smoothing function in the search algorithm versus the bin width of our single-pulse profiles used for the pulse amplitude plots). The relationship between S/N and time resolution is given in \citet{kp15}.

\begin{figure}[t]
\vspace*{3.8cm}
\centering 
\includegraphics[trim=100 250 80 250, width=9.5cm]{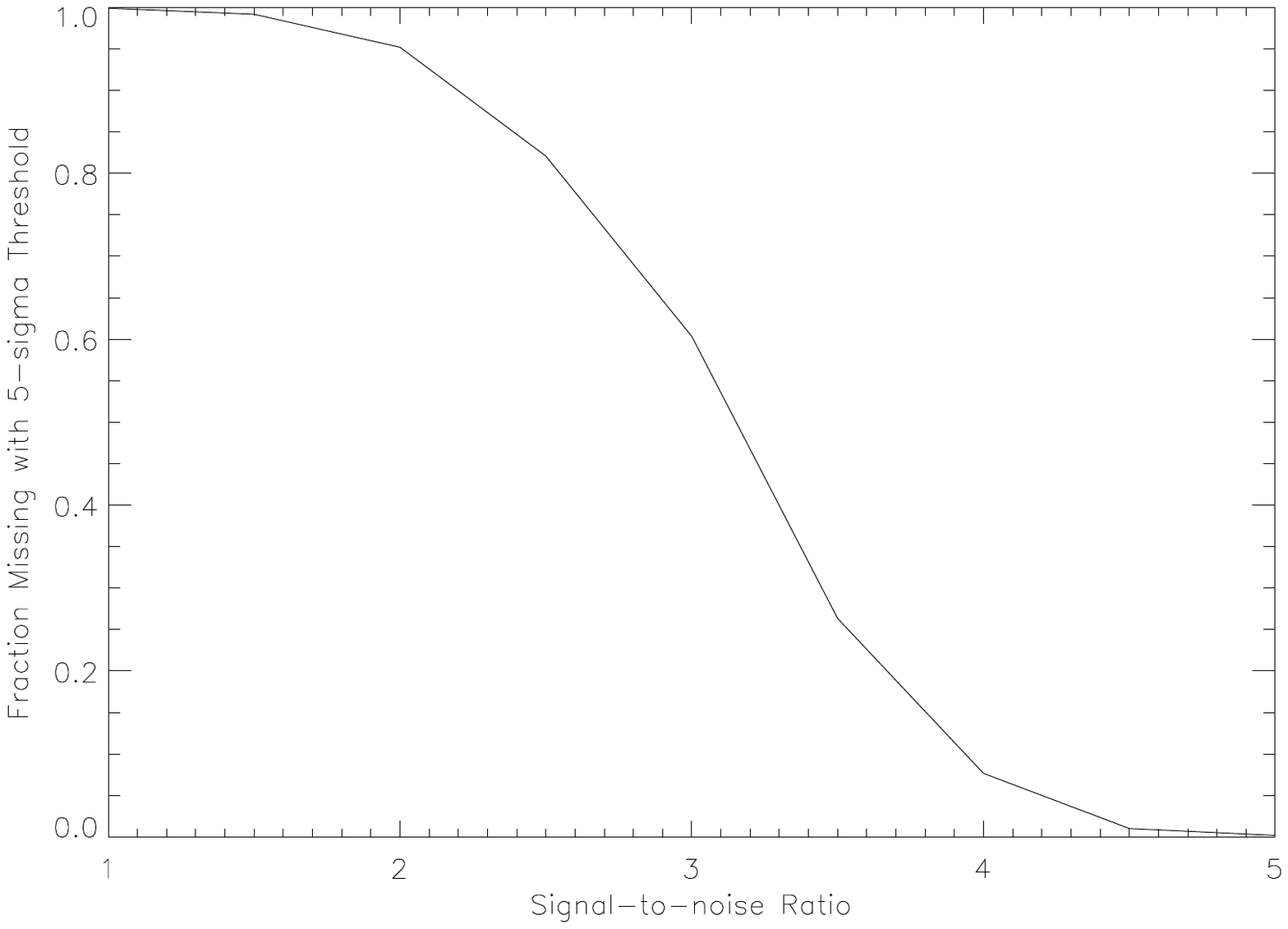}
\vspace*{-2.2cm}
\caption{Fraction of pulses missed vs. S/Ns in our detectability simulations. The period in the simulated timeseries is 1 s. For S/Ns larger than four, the missed fraction is small, which means most of the pulses should be detected despite other factors such as human error in the process. Note that this is not the S/N returned by the single-pulse search code (discussed in section \ref{ssec:5.1}), and the simulation parameters such as period, sample time, and pulse width must be specified for different RRATs.} 
\label{fig_det}
\end{figure} 

\begin{figure}[t]
\vspace*{0cm}
\centering 
\captionsetup{justification=centering}
\vspace*{-0.9cm}%
 \begin{minipage}[b]{0.45\textwidth}%
\centering \includegraphics[trim=100 250 10 0, width=6.5cm]{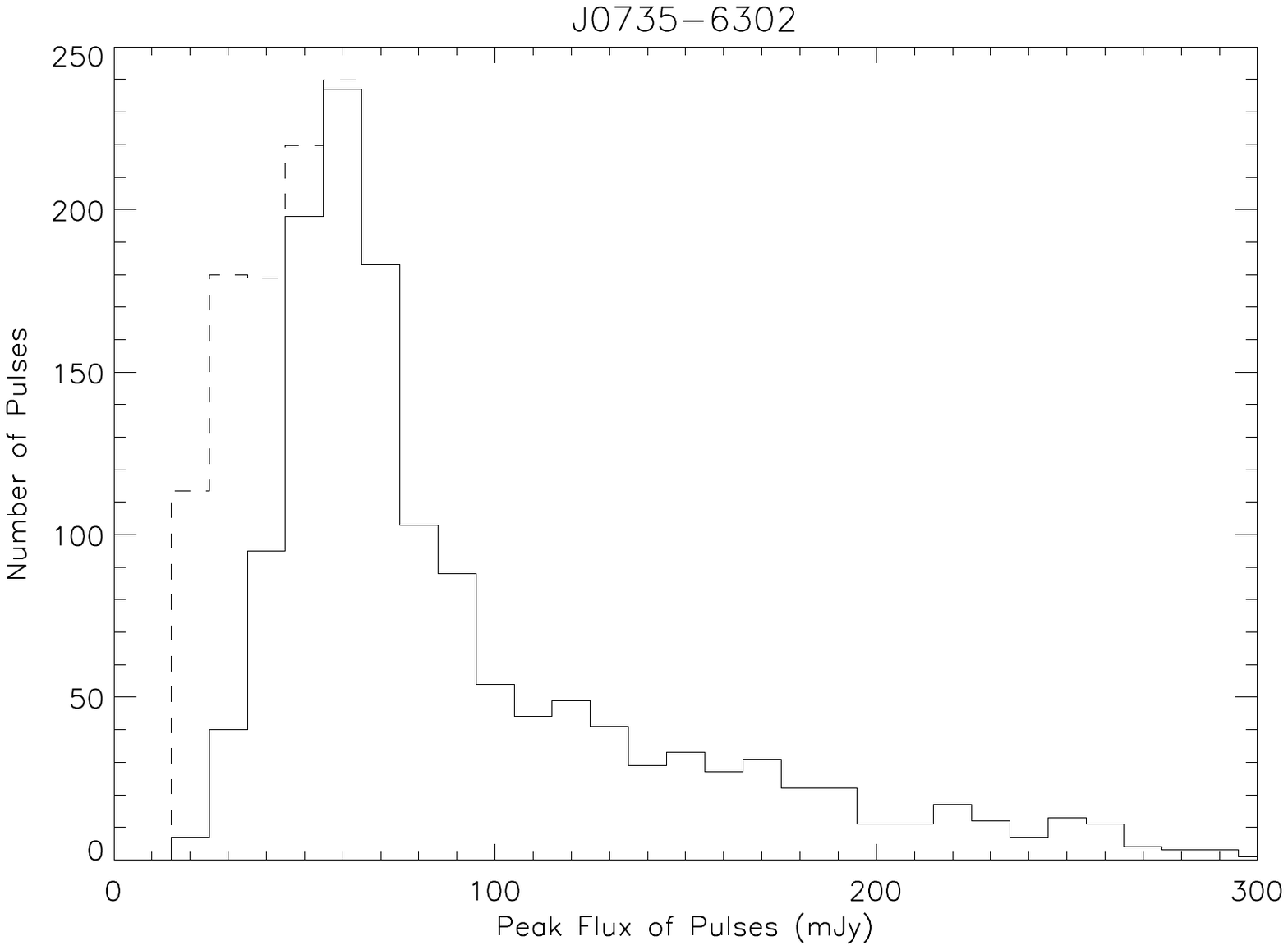}
\end{minipage}%
 \begin{minipage}[b]{0.45\textwidth}%
\centering \includegraphics[trim=100 250 10 0, width=6.5cm]{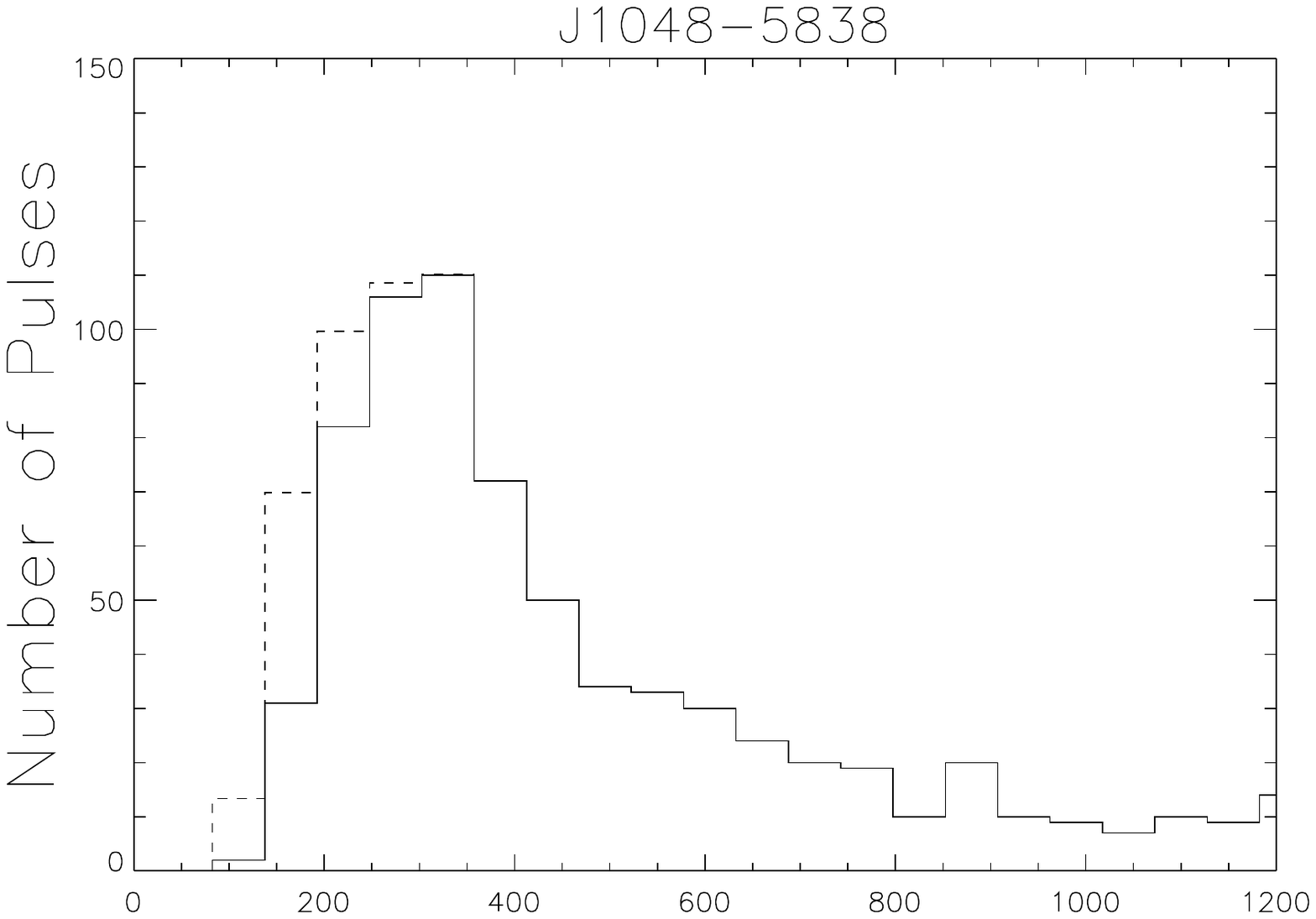}
 \end{minipage}
\vfill
\vspace*{-2.1cm}
 \begin{minipage}[b]{0.45\textwidth}%
\centering \includegraphics[trim=100 250 10 0, width=6.5cm]{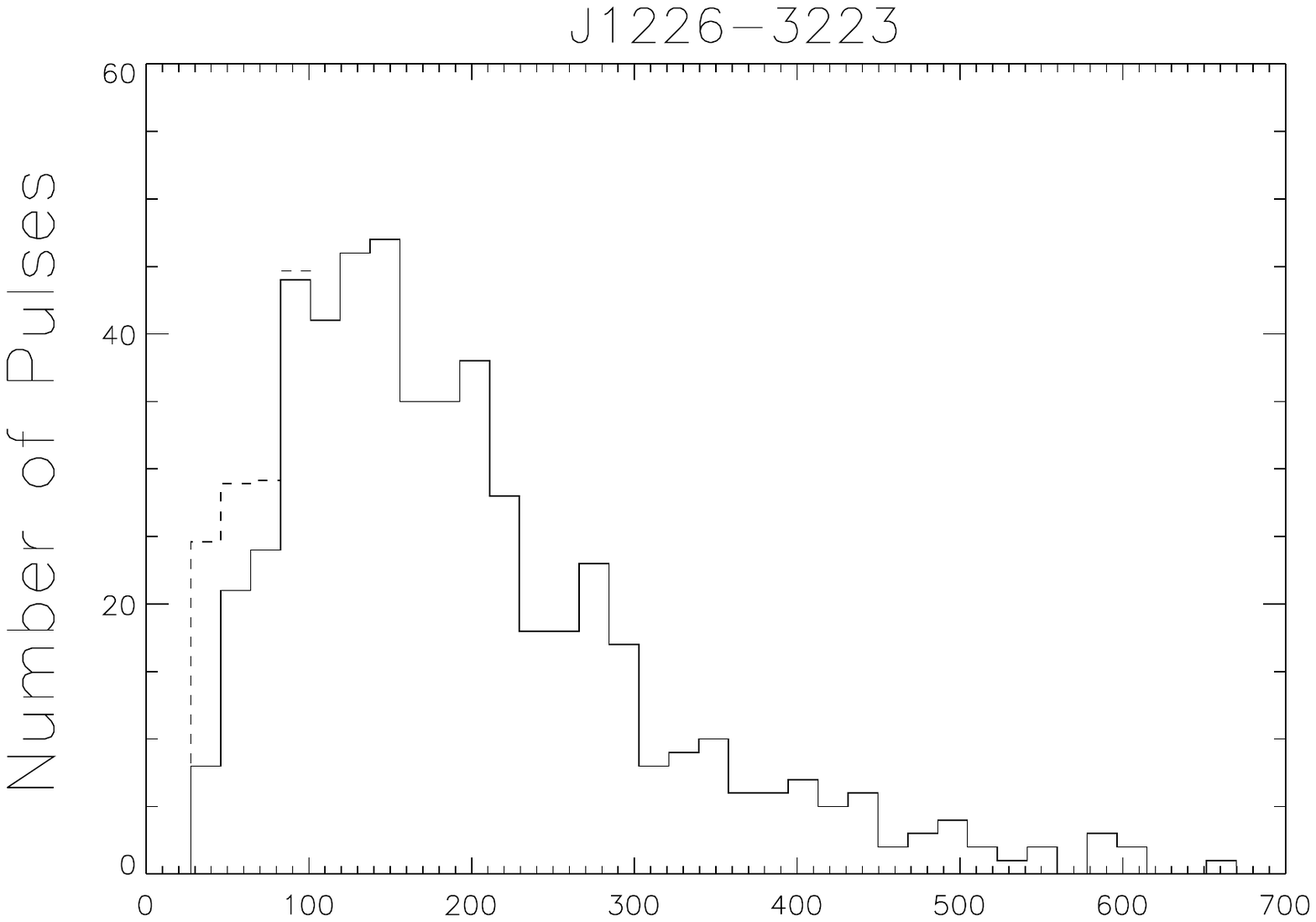}
 \end{minipage}
 \begin{minipage}[b]{0.45\textwidth}%
\centering \includegraphics[trim=100 250 10 0, width=6.5cm]{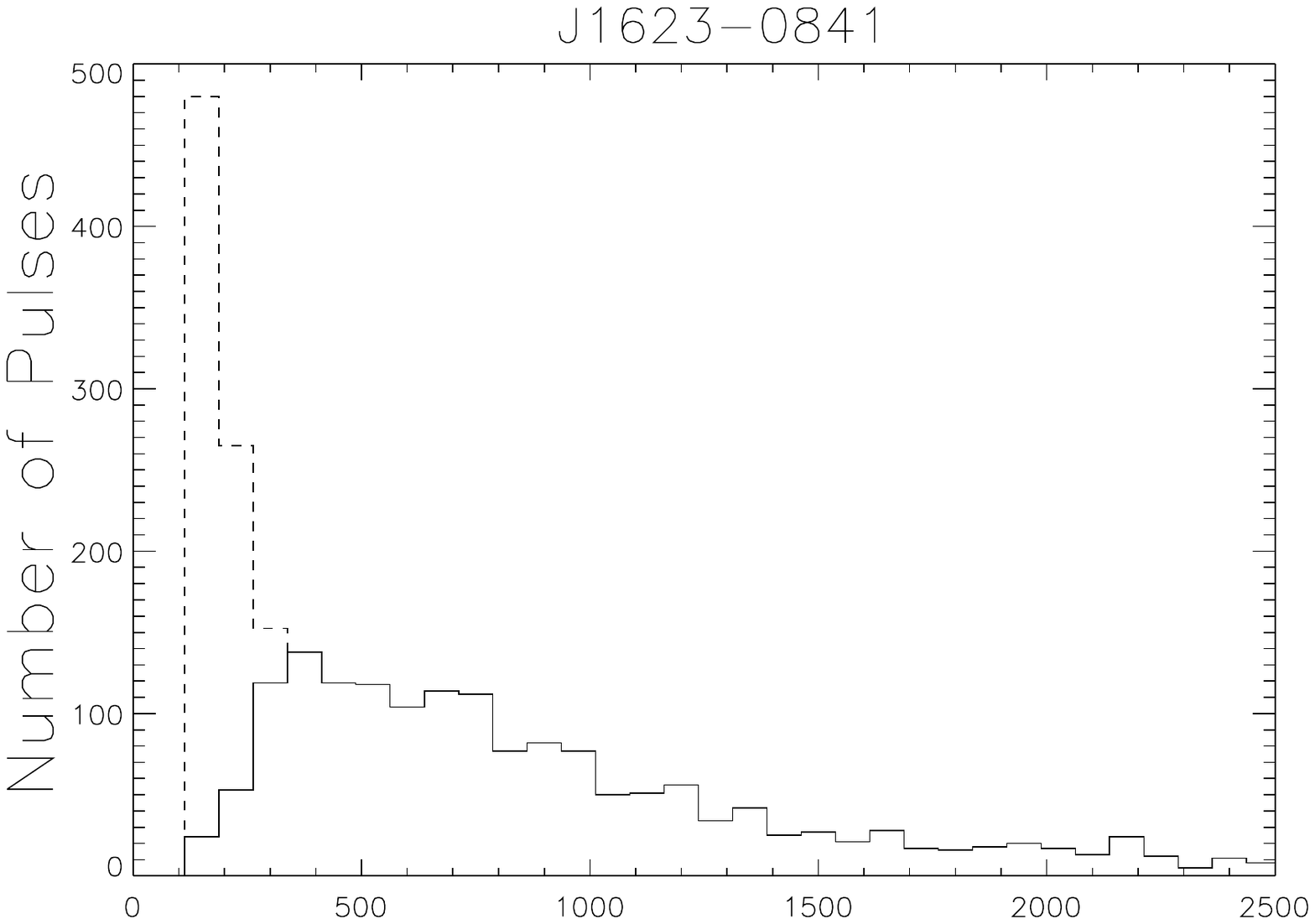}
 \end{minipage}
\vfill 
\vspace*{-2.1cm}
 \begin{minipage}[b]{0.45\textwidth}%
\centering \includegraphics[trim=100 250 10 0, width=6.5cm]{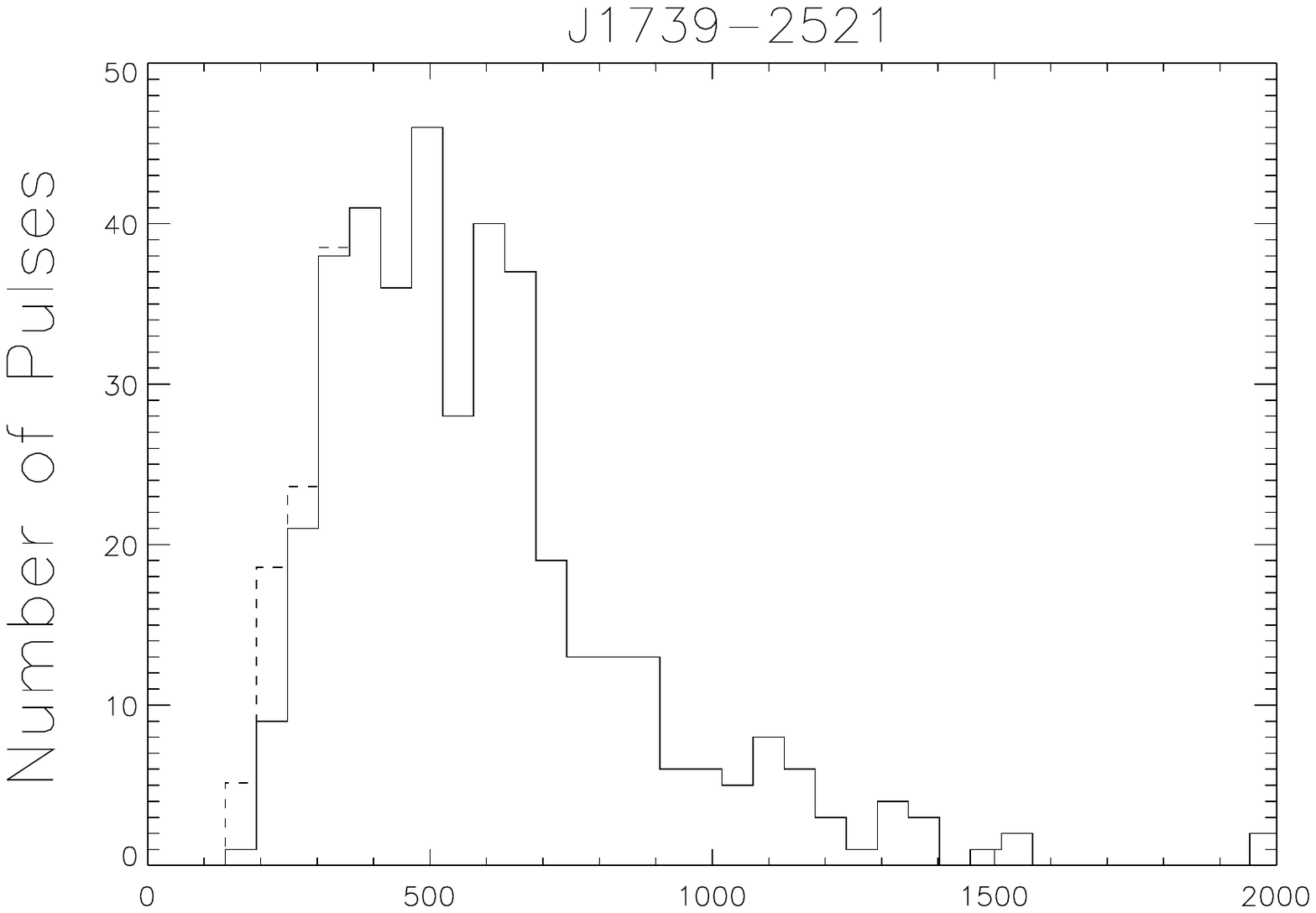}
 \end{minipage}%
 \begin{minipage}[b]{0.45\textwidth}%
\centering \includegraphics[trim=100 250 10 0, width=6.5cm]{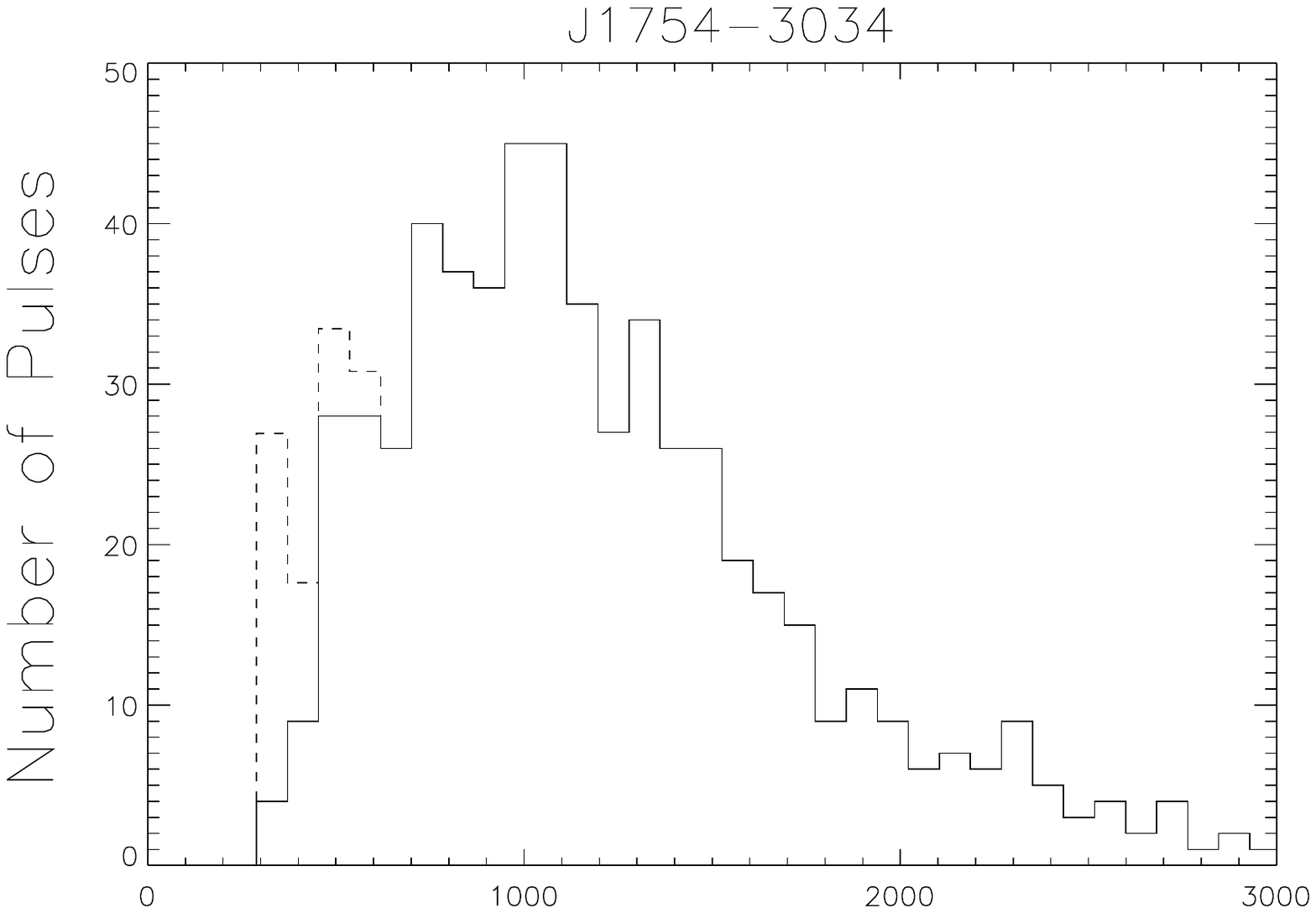}
 \end{minipage}%
\vfill
\vspace*{-1.2cm}
 \begin{minipage}[b]{0.45\textwidth}%
\centering \includegraphics[trim=70 20 0 0, width=5.4cm]{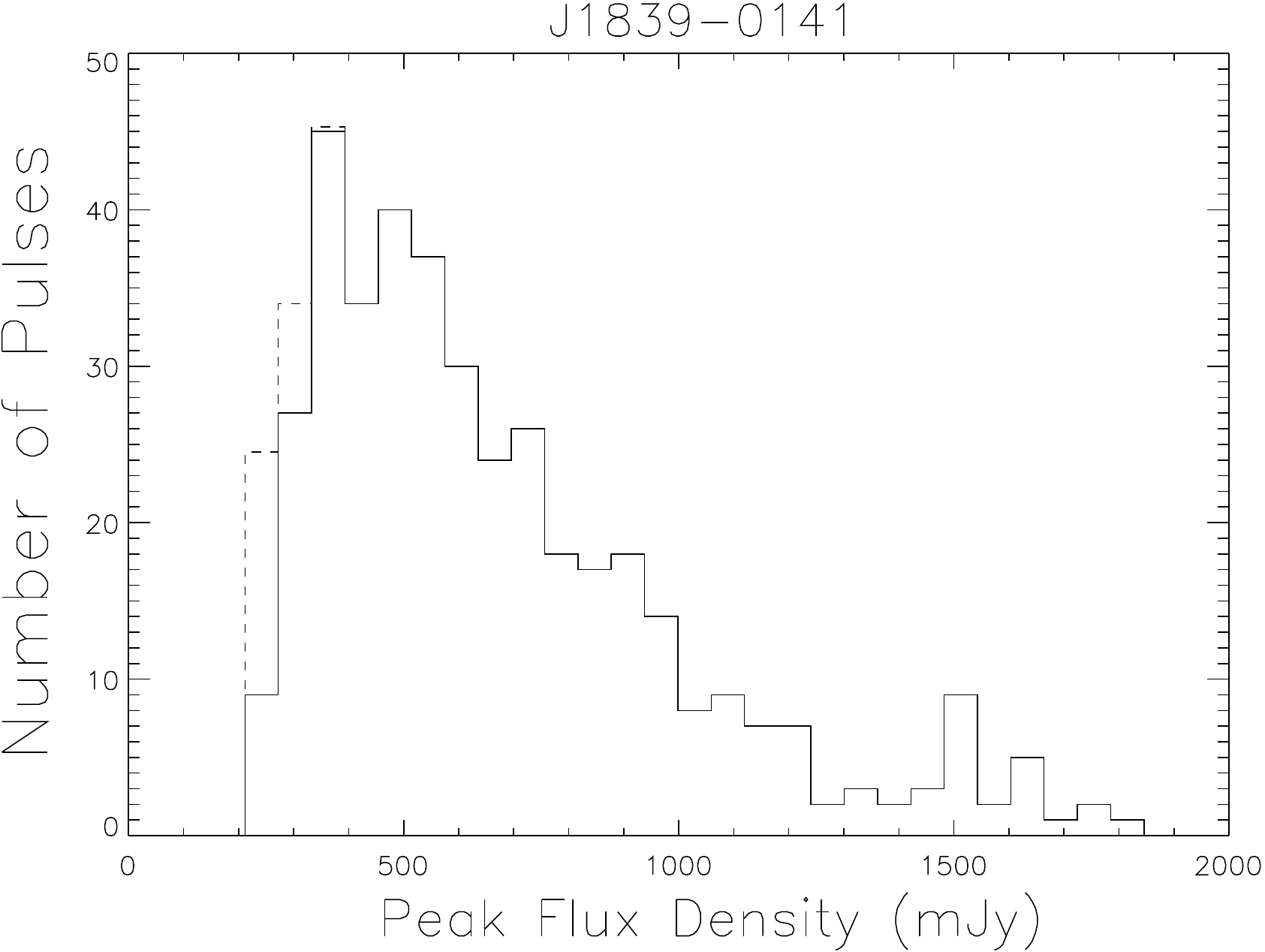}
 \end{minipage}%
 \begin{minipage}[b]{0.45\textwidth}%
\centering \includegraphics[trim=75 20 0 0, width=5.4cm]{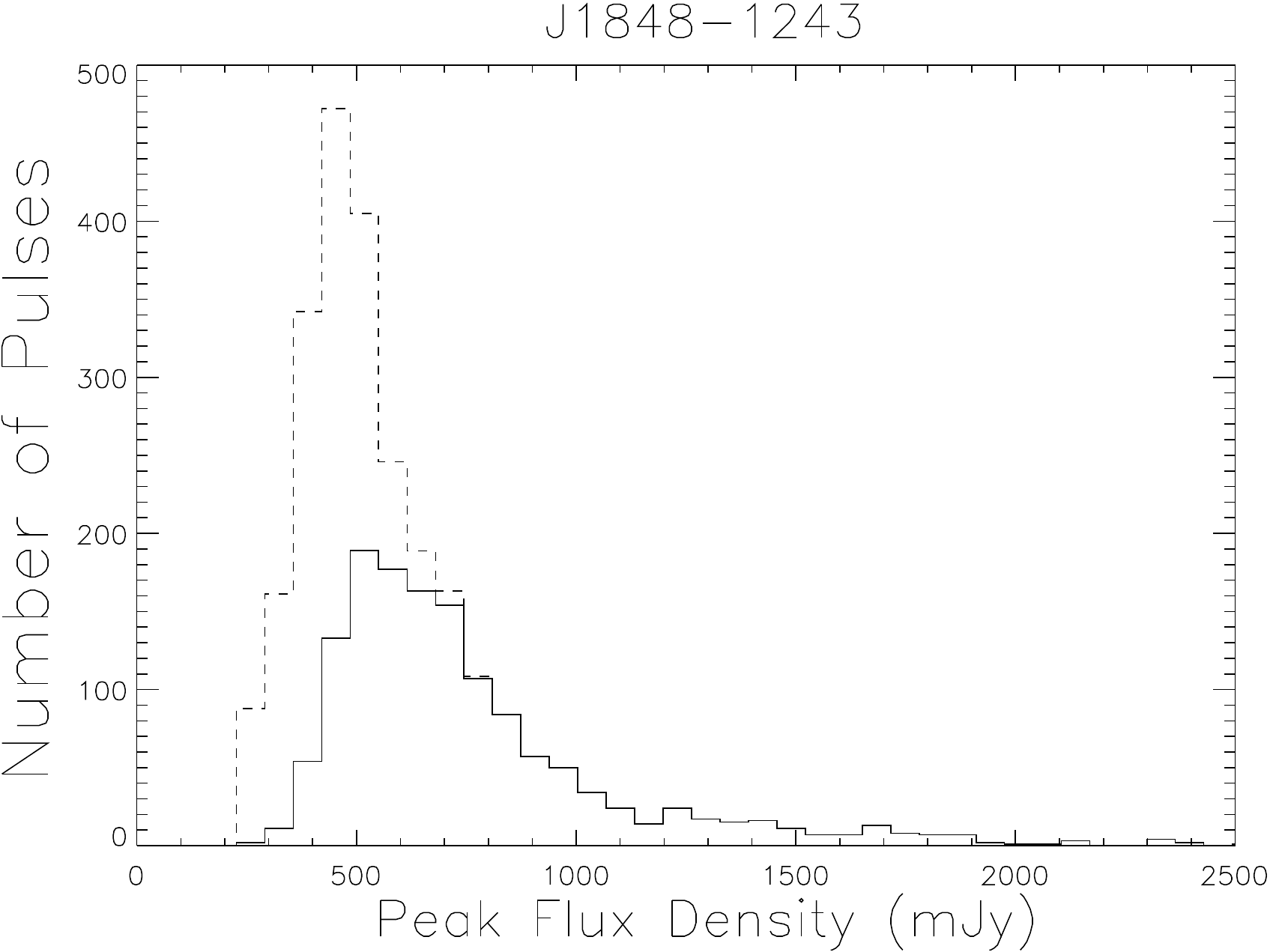}
 \end{minipage}%
\vspace*{0.4cm}
\caption{The detectability correction to the pulse S/N distribution for eight RRATs. The dotted lines are the expected ``real'' S/N distributions after correction for all pulses from the RRATs above detection threshold, and the solid lines are the original S/N distributions before correction for pulses detected. From this figure we can see that many weak pulses are missed in the single-pulse detection algorithm for PSRs J0735$-$6302, J1623$-$0841, and J1848$-$1243, which was seen in Figure \ref{fig4}.} 
\label{fig_real}
\vspace*{-0.4cm}
\end{figure}

We simulated several timeseries with different peak profile S/Ns varying from one to five but with the same period, dm, pulse width, and sample time corresponding to a specific RRAT. With the commonly used detection threshold S/N = 5 (which is calculated as a search S/N), the result of an example test is shown in Figure \ref{fig_det}. It is clear that only pulses with profile S/Ns larger than four are reliably detected, and the numbers of pulses with lower S/Ns are obviously underestimated. We therefore correct the amplitude distributions for this effect for these eight RRATs, shown in Figure \ref{fig_real}. The dotted line shows the expected ``real'' S/N distribution after the correction to the detected pulse distribution (solid line). These corrections vary for different RRATs due to their period, sample time, and pulse width. We can clearly see some of them indicate a considerable number of missed pulses. In the mean time, the result shows that those RRATs being detected with large numbers of pulses (J1623$-$0841 and J1848$-$1243) also have large fractions of missed pulses. This may indicate that they are not truly sporadic and we are detecting only part of their pulse distributions. 

\subsection{Weak Pulse Analysis}

Inspecting the pulse amplitude distributions shown in Figure \ref{fig_real}, it is true that there are a fair number of weak pulses with S/Ns lower than the threshold that are not detectable by the single-pulse search pipeline. In order to confirm this, we constructed `weak pulse' profiles by subtracting all detected single-pulse profiles from the composite profile created through folding all data in each observation using the timing model. The profiles here are scaled with numbers of the pulses per observation so that each observation has the same weight. The results are shown in the lower panels of Figure \ref{fig4} for all of the RRATs. The weak pulse profiles of PSR J1048$-$5838 and J1839$-$0141 show no significant emission. On the other hand, there is a significant peak in the weak pulse profile for all other RRATs, especially J0735$-$6302, J1623$-$0841, and J1848$-$1243, which indicates that a large number of weak pulses were missed in the search. Interestingly, PSR J1226$-$3223 has a somewhat significant peak but with a small number of missed pulses. This is probably attributed to the low mean pulse flux density and the low DM, which caused a larger fraction of pulses to remain undetected due to low S/N or RFI effects. These results can also provide a test of the hypothesis that RRATs are normal pulsars, with some pulses not detectable due to distance \citep{wsrw06}. For the RRATs that we can see, no corresponding peak in these weak pulse profiles, e.g. J1048$-$5838 and J1739$-$2521, thus such an effect is ruled out.

\subsection{Periodicities in the Emission Timescales} 

\begin{figure}[t]
\vspace*{2.0cm}
\centering
\includegraphics[trim=170 490 180 240, width=7cm]{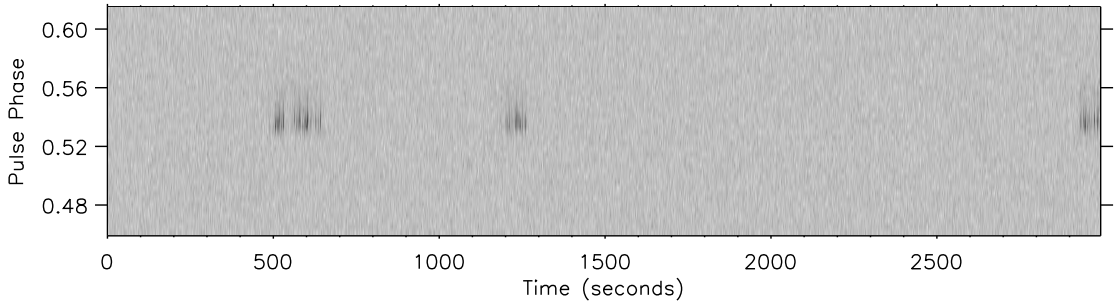}
\caption{Pulsar folding plot shows ``on'' and ``off'' phases for PSR J1839$-$0141. In this figure, we plot pulsar signal in pulse phase versus time, where vertical short dark lines indicate emission from the pulsar. We can see that this RRAT repeatedly turns on and off three times in one single observation with each ``on'' phase lasting for one or two minutes.} 
\label{fig-on}
\end{figure}

\begin{figure}[!htb]
\centering
\captionsetup{justification=centering}
\vspace*{-1.3cm}%
\hspace{0.5cm}
 \begin{minipage}[b]{0.45\textwidth}%
\centering 
\includegraphics[trim=100 520 10 0, width=7.5cm, height=3.8cm]{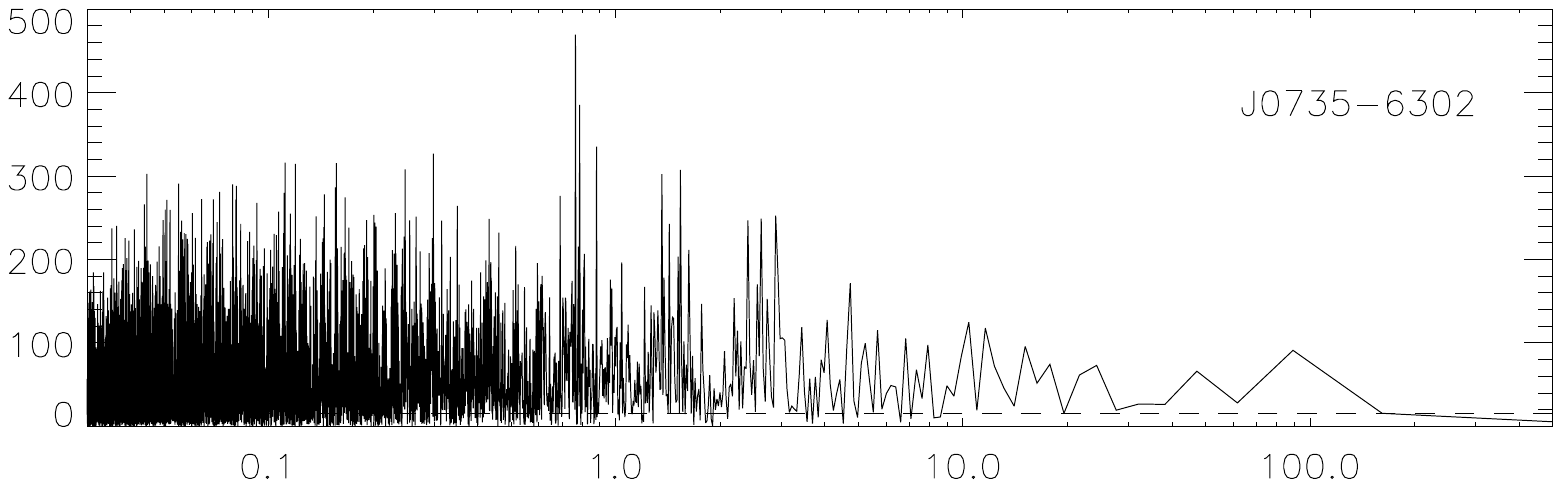}
\end{minipage}%
\hspace*{0.5cm}
 \begin{minipage}[b]{0.45\textwidth}%
\centering \includegraphics[trim=100 520 10 0, width=7.5cm, height=3.8cm]{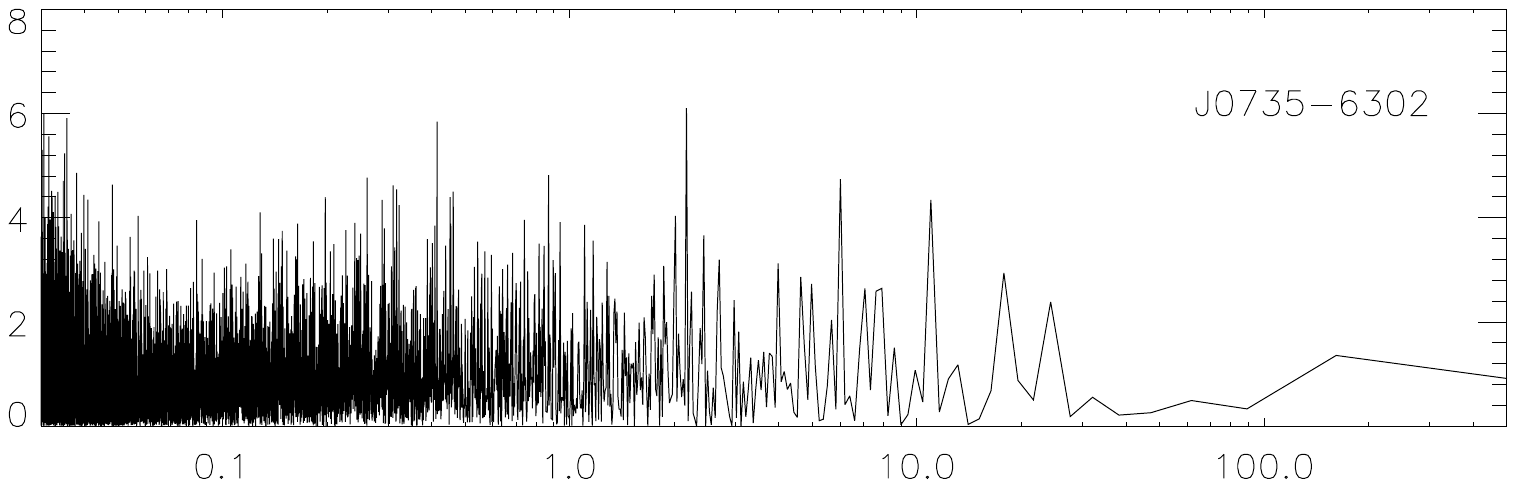}
 \end{minipage}
\vfill
\vspace*{-1.9cm}
\hspace{0.5cm}
 \begin{minipage}[b]{0.45\textwidth}%
\centering \includegraphics[trim=100 520 10 0, width=7.5cm, height=3.8cm]{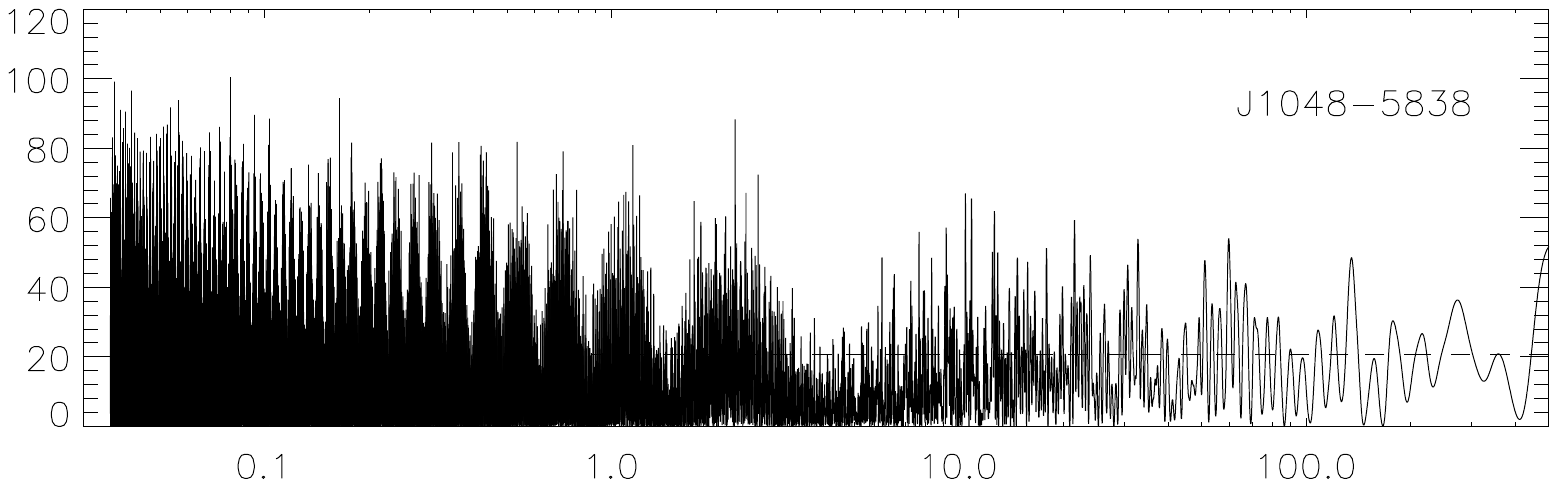}
 \end{minipage}%
\hspace*{0.5cm}
 \begin{minipage}[b]{0.45\textwidth}%
\centering \includegraphics[trim=100 520 10 0, width=7.5cm, height=3.8cm]{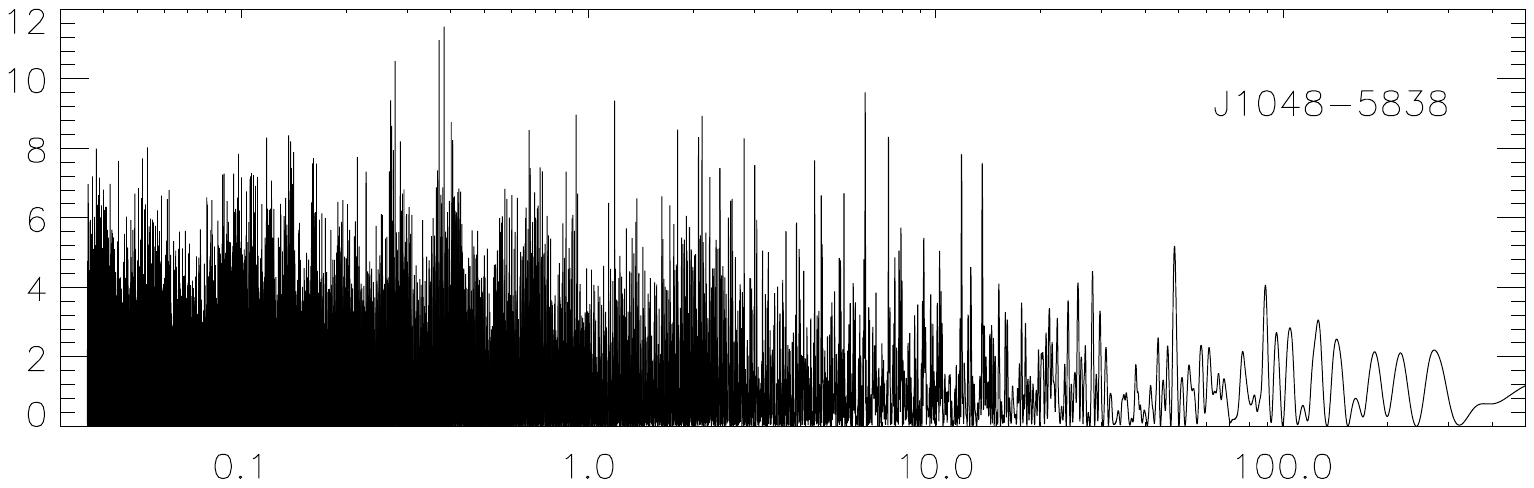}
 \end{minipage}
\vfill 
\vspace*{-1.9cm}
\hspace{0.5cm}
 \begin{minipage}[b]{0.45\textwidth}%
\centering \includegraphics[trim=100 520 10 0, width=7.5cm, height=3.8cm]{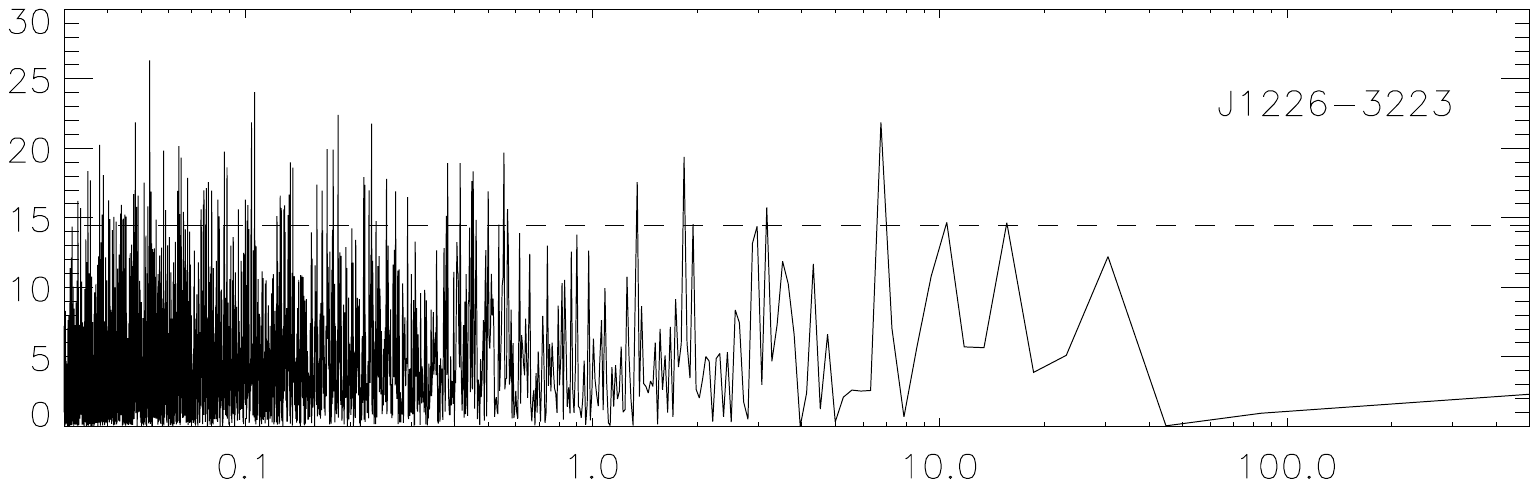}
 \end{minipage}%
\hspace*{0.5cm}
 \begin{minipage}[b]{0.45\textwidth}%
\centering \includegraphics[trim=100 520 10 0, width=7.5cm, height=3.8cm]{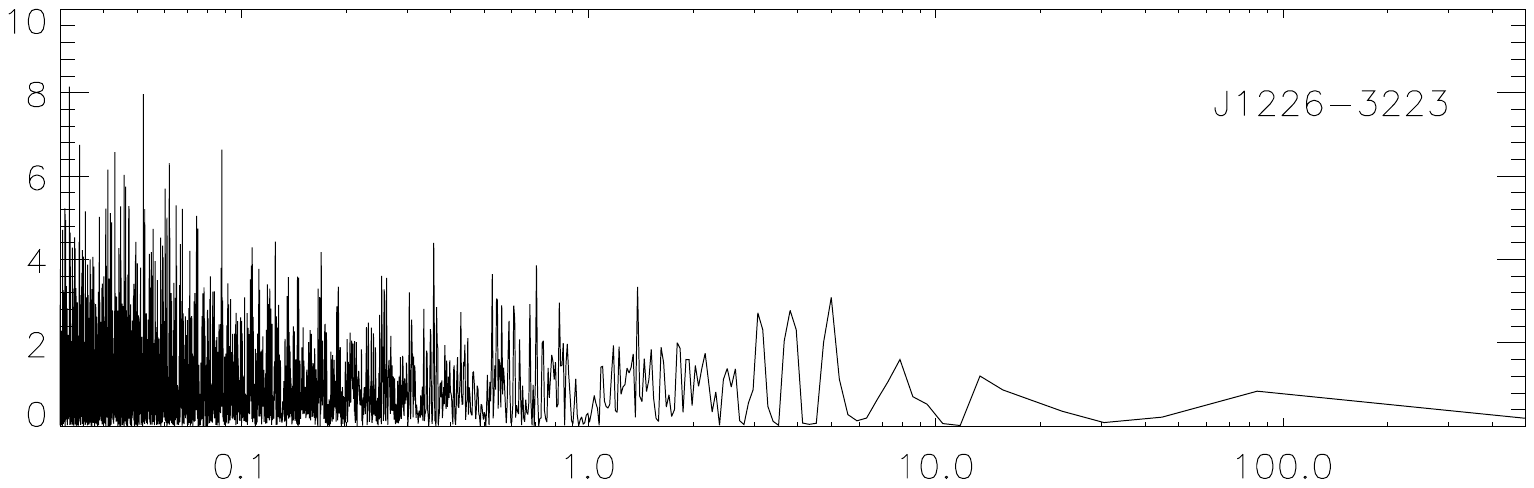}
 \end{minipage}
\vfill 
\vspace*{-1.9cm}
\hspace{0.5cm}
 \begin{minipage}[b]{0.45\textwidth}%
\centering \includegraphics[trim=100 520 10 0, width=7.5cm, height=3.8cm]{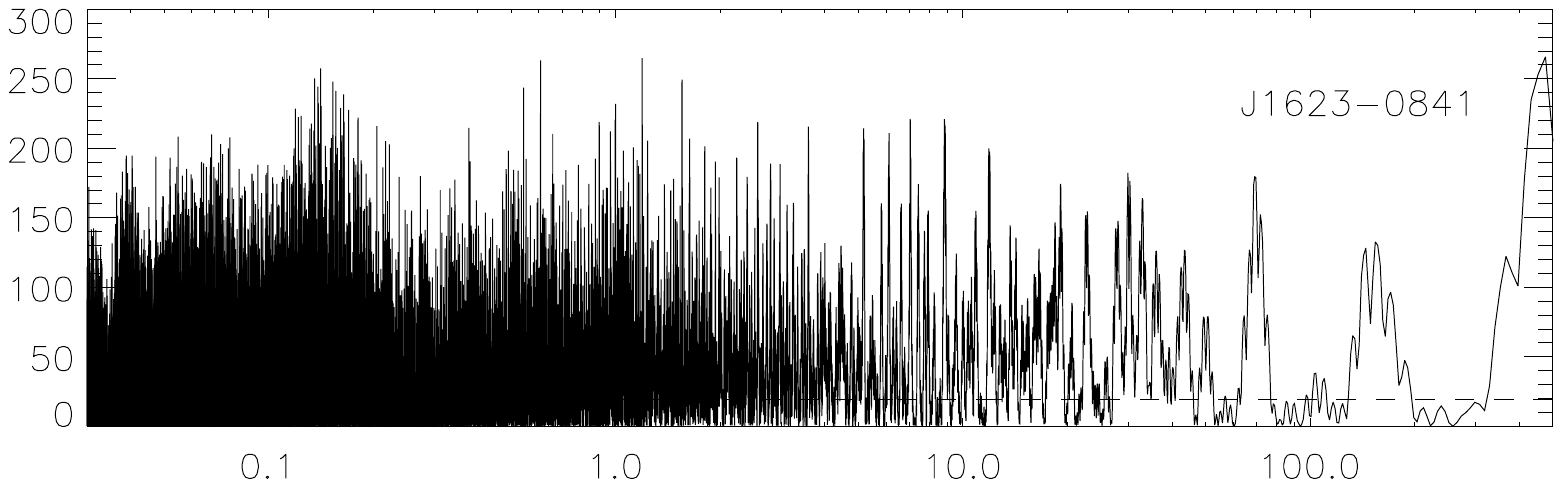}
 \end{minipage}%
\hspace*{0.5cm}
 \begin{minipage}[b]{0.45\textwidth}%
\centering \includegraphics[trim=100 520 10 0, width=7.5cm, height=3.8cm]{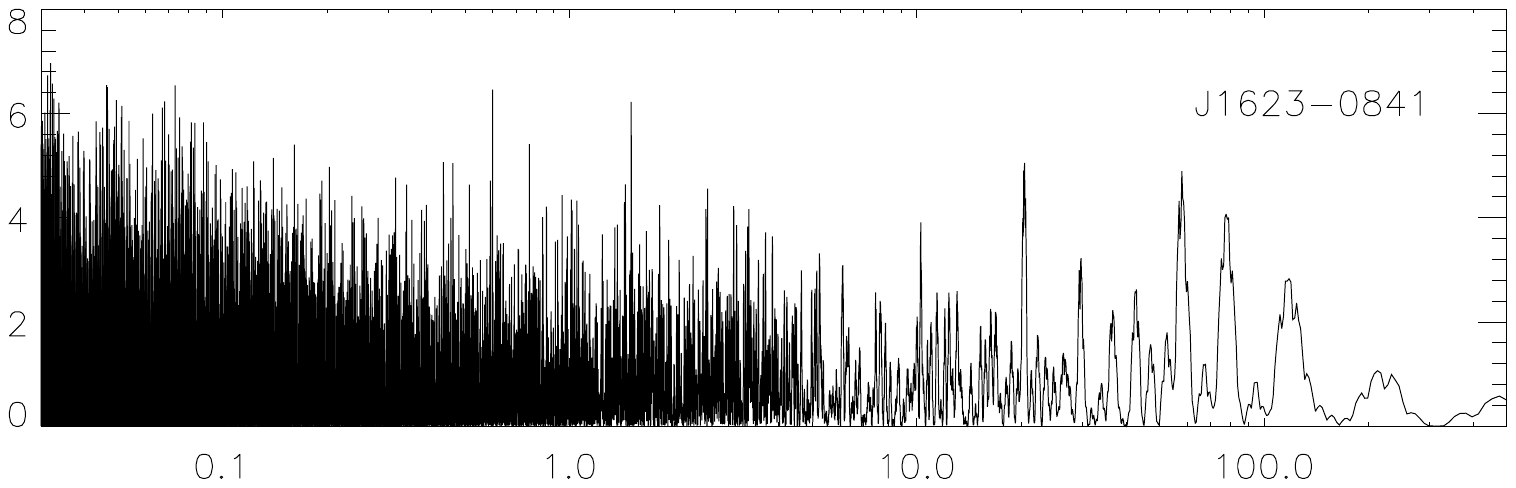}
 \end{minipage}
\vfill 
\vspace*{-1.9cm}
\hspace{0.5cm}
 \begin{minipage}[b]{0.45\textwidth}%
\centering \includegraphics[trim=100 520 10 0, width=7.5cm, height=3.8cm]{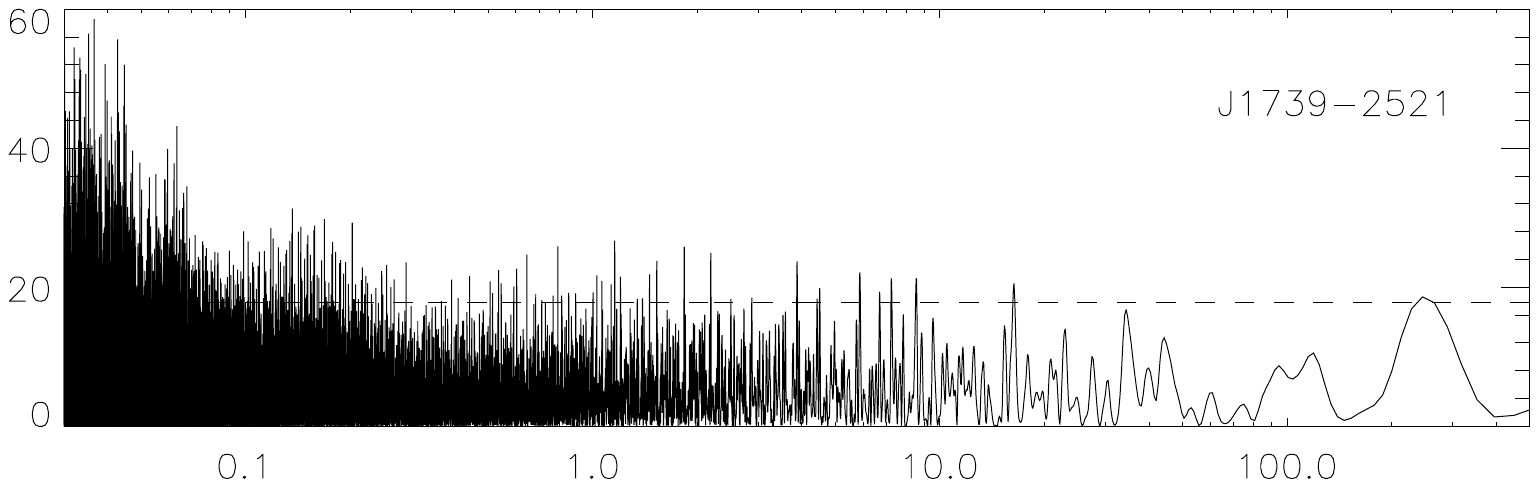}
 \end{minipage}%
\hspace*{0.5cm}
 \begin{minipage}[b]{0.45\textwidth}%
\centering \includegraphics[trim=100 520 10 0, width=7.5cm, height=3.8cm]{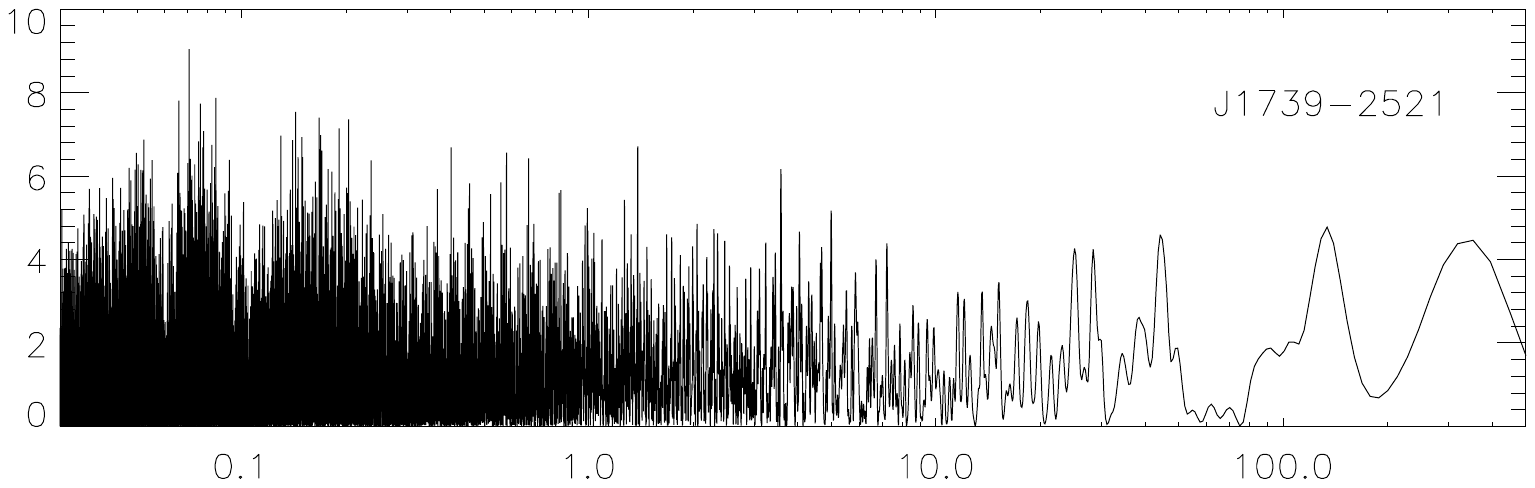}
 \end{minipage}
\vfill 
\vspace*{-1.9cm}
\hspace{0.5cm}
 \begin{minipage}[b]{0.45\textwidth}%
\centering \includegraphics[trim=100 520 10 0, width=7.5cm, height=3.8cm]{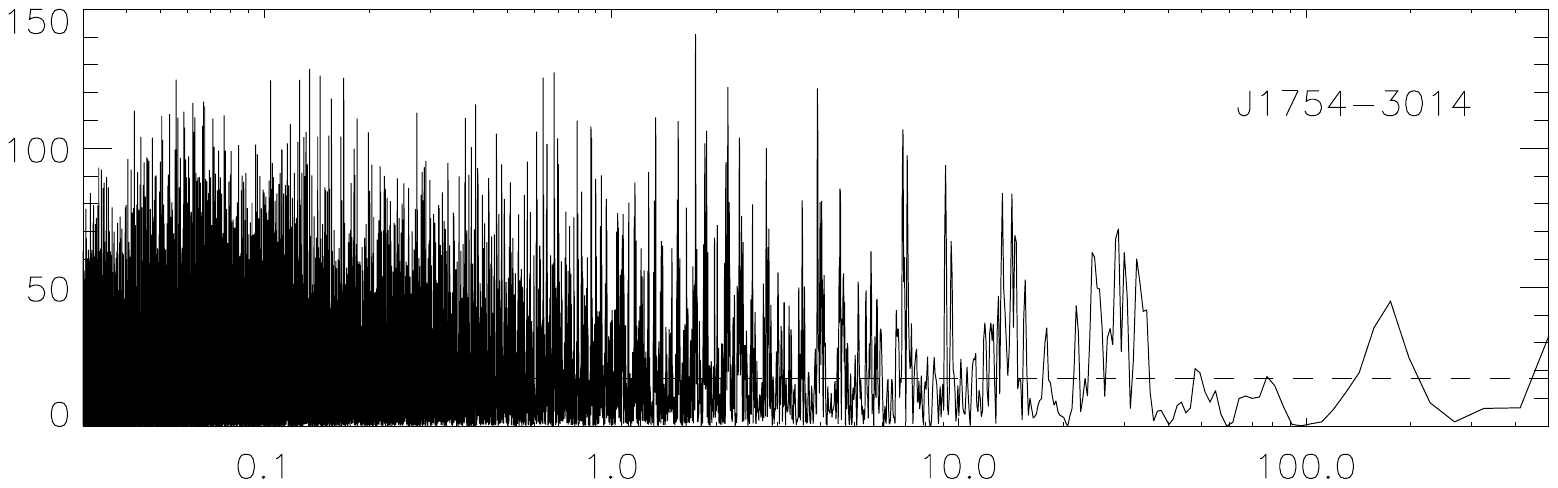}
 \end{minipage}%
\hspace*{0.5cm}
 \begin{minipage}[b]{0.45\textwidth}%
\centering \includegraphics[trim=100 520 10 0, width=7.5cm, height=3.8cm]{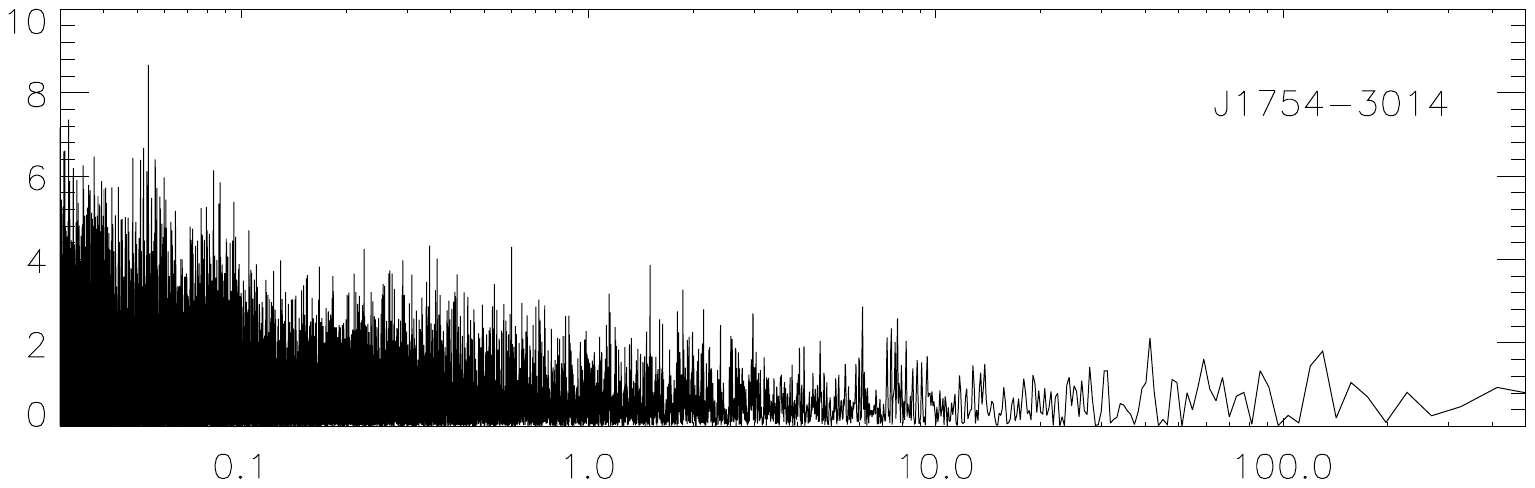}
 \end{minipage}
\vfill 
\vspace*{-1.9cm}
\hspace{0.5cm}
 \begin{minipage}[b]{0.45\textwidth}%
\centering \includegraphics[trim=100 520 10 0, width=7.5cm, height=3.8cm]{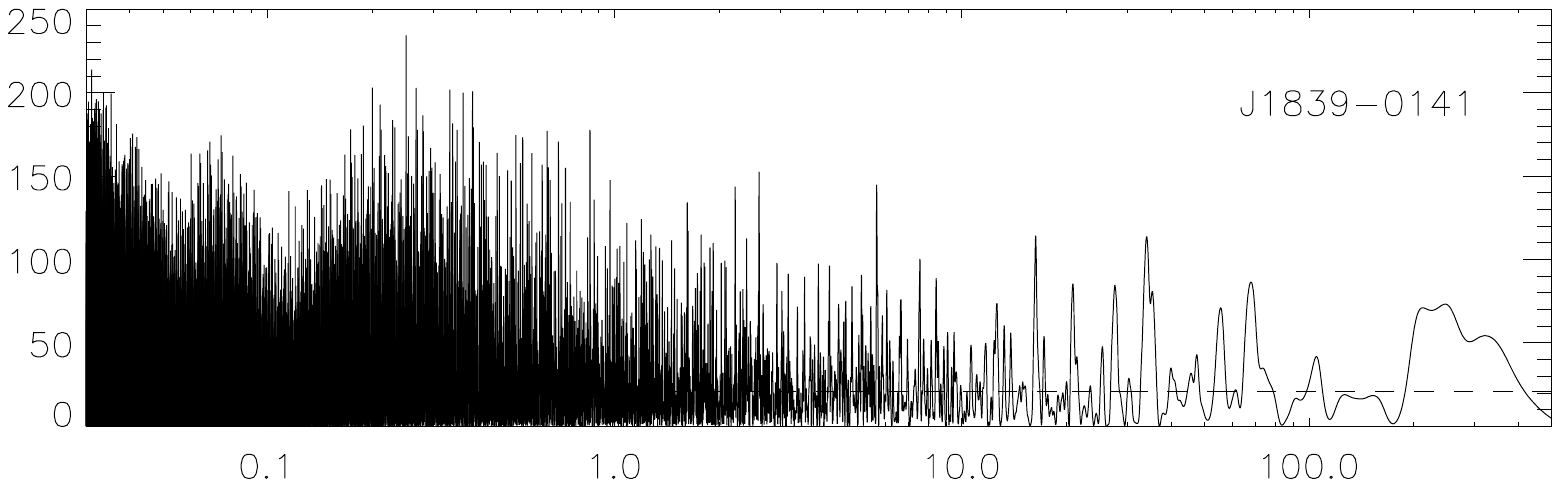}
 \end{minipage}%
\hspace*{0.5cm}
 \begin{minipage}[b]{0.45\textwidth}%
\centering \includegraphics[trim=100 520 10 0, width=7.5cm, height=3.8cm]{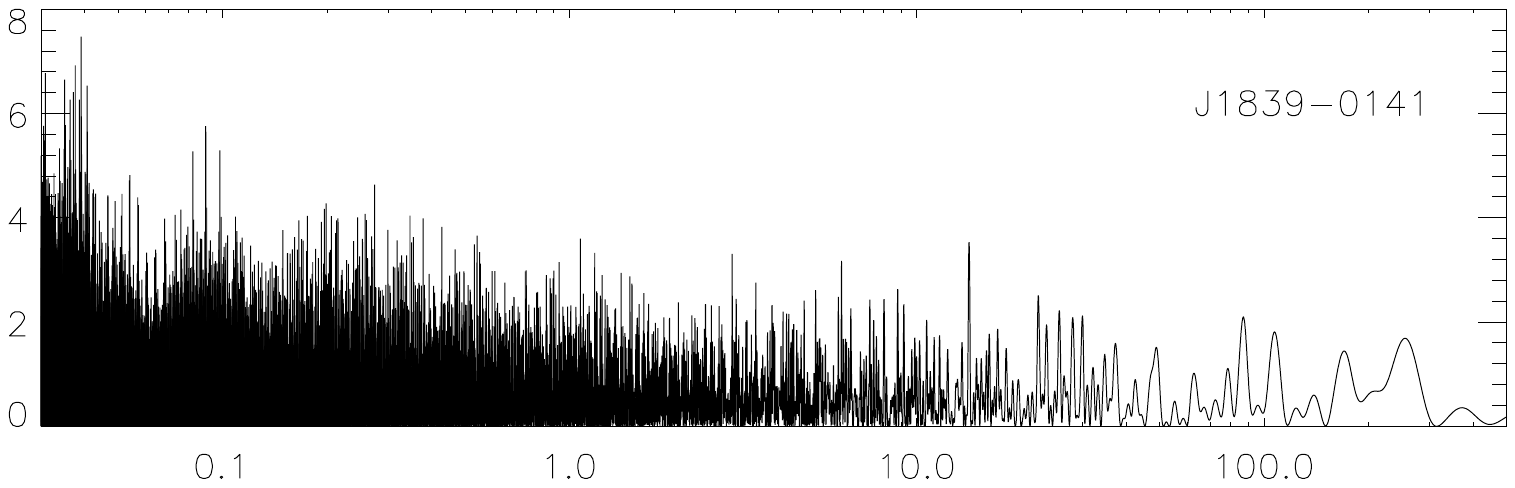}
 \end{minipage}
\vfill 
\vspace*{-1.9cm}
\hspace{0.5cm}
 \begin{minipage}[b]{0.45\textwidth}%
\centering \includegraphics[trim=100 520 10 0, width=7.5cm, height=3.8cm]{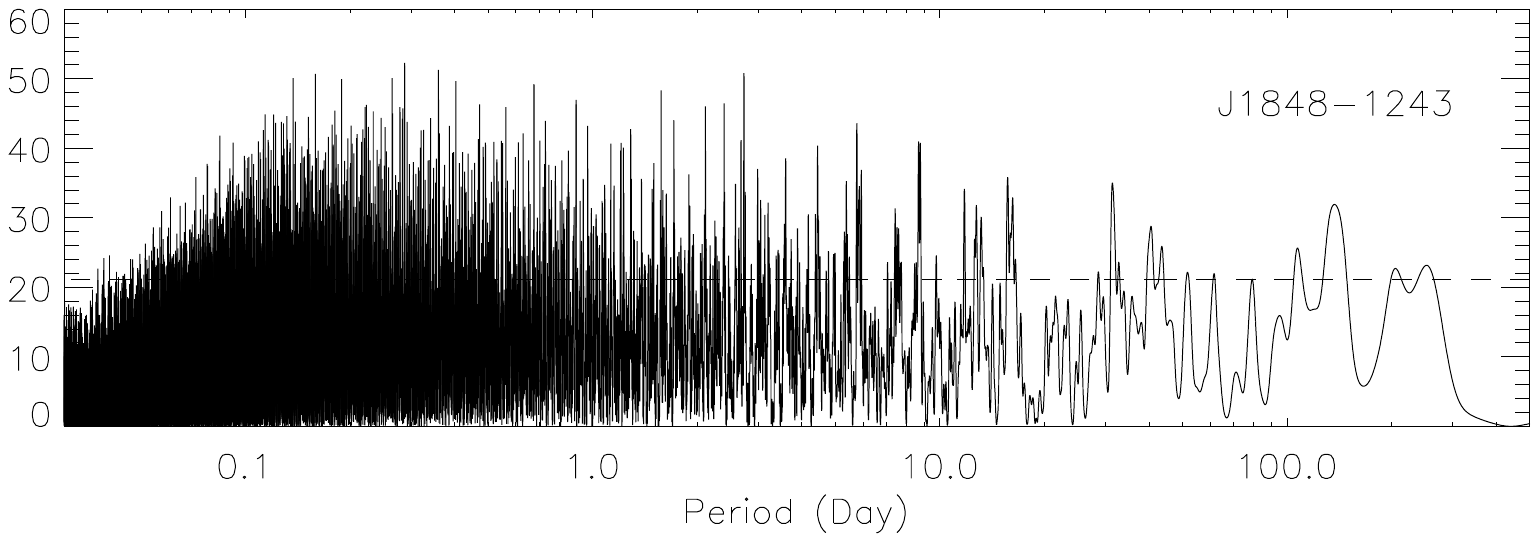}
 \end{minipage}%
\hspace*{0.5cm}
 \begin{minipage}[b]{0.45\textwidth}%
\centering \includegraphics[trim=100 520 10 0, width=7.5cm, height=3.8cm]{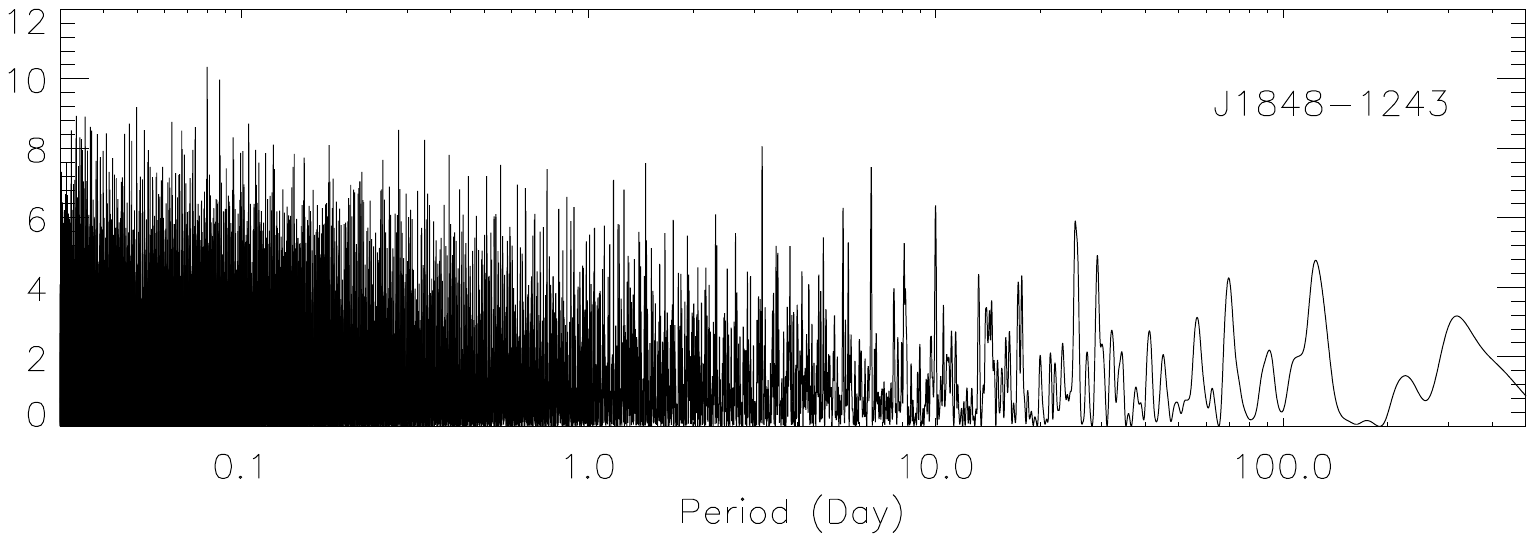}
 \end{minipage}
\vfill 
\vspace*{-20.48cm}
\hspace*{1.02cm}
\includegraphics[trim=100 500 150 025, height=14.11cm, width=21.75cm]{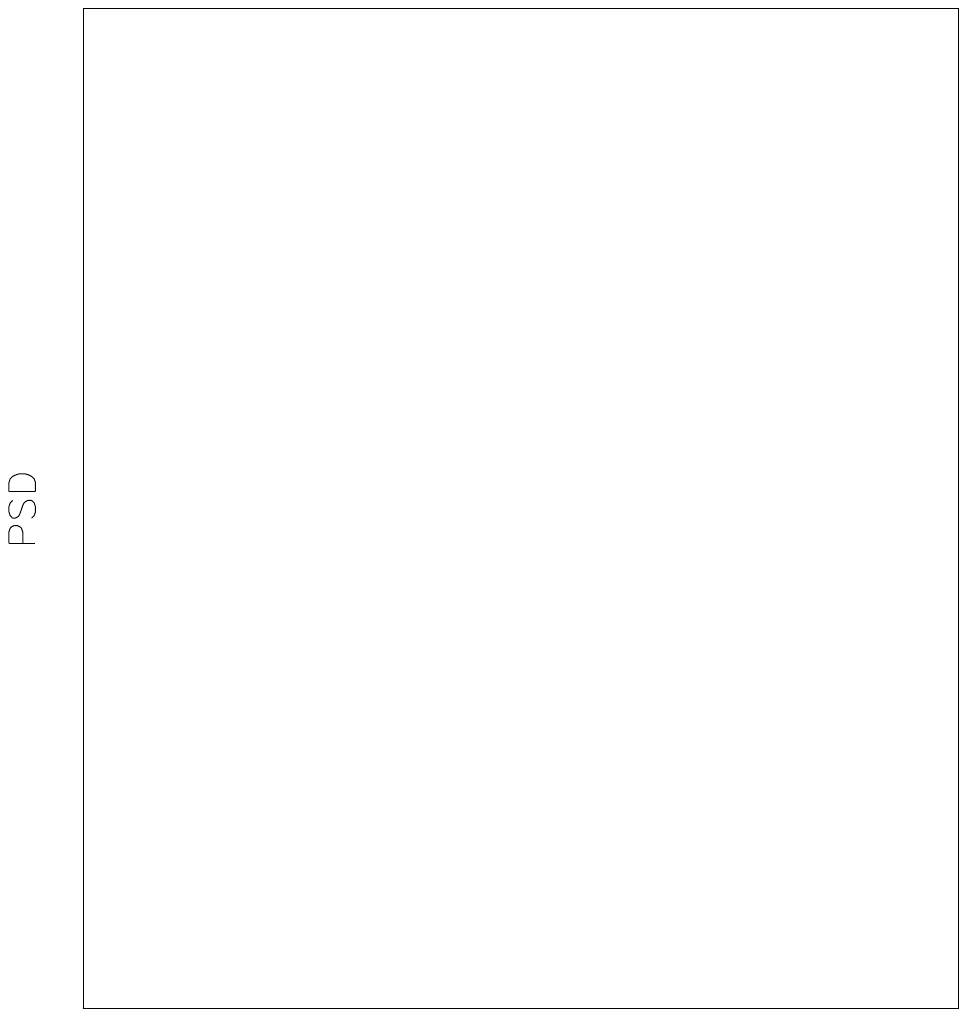}
\vspace*{5.5cm}
\caption{Lomb$-$Scargle test results for all eight RRATs. Here the power spectral densities (PSDs) are plotted versus log-scaled periods of emission in units of days. The peaks in the PSD curves show the most significant candidates, while the dashed line provides a 2.5$\sigma$ (99\%) significance threshold. The plots in the left column show tests on the original RRATs pulse arrival times, and in the right are the tests upon randomized time series with same spin period and observation time as on the left. The 2.5$\sigma$ dashed line is not visible because it lies above all of the randomized data points. Note the difference in scales of PSD axis between the real data and randomized data. We can see that, for all our RRATs, we detect PSD peaks that exceed the threshold and are far higher than the random test, indicating that all of these RRATs show true periodic behavior in their emission times.}
\label{fig6}
\end{figure}

As shows in Figure \ref{fig-on}, some RRATs turn ``on'' and ``off'' seemingly regularly, so there is a possibility that some periodicities exist in their on/off timescales. We have applied Lomb$-$Scargle analysis \citep{sca82} to the unevenly sampled pulse arrival times for this task. These time series data include all detected pulse arrival times and the times of rotations without a detected pulse within the time span of all radio observations. This L$-$S analysis can then reveal possible periodicities with significances proportional by the power spectral density. Given the same numbers of pulses but randomly varying their position in time series by shuffling their flags of ``on'' and ``off'' for every arrival time. Keeping the same windows of observation and gaps between the observations, we also performed the same analysis in order to check whether the observed pulse time sequences are consistent with a random distribution. Details of this technique can be found in \citet{pmk+11}. Figure \ref{fig6} shows the spectra for the actual data and an example of one randomized time series. All spectra have peaks of over 2.5$\sigma$ (99\%) significance. In order to test whether this result is reasonable, we created 1000 random realizations by randomly placing each detected pulse for each RRAT with the observation windows. No peak was detected at the same or higher significance level detected within several bins at the discovered peak frequencies, or in the overall frequencies, in any of these realizations. This is consistent with the derived more than 99\% significance of these periodicities. Periodicities that are longer than half of the overall observation time span are ignored. Examining these frequencies with significant power, we can say with confidence that emission periodicity exists at different timescales for all eight RRATs checked. The periodicities with the largest significance are listed in Table \ref{tab3} in order of their significance level.

\begin{table}[!t]
\vspace{0.2cm}
\end{table}

\floattable

\begin{deluxetable*}{lccccccr}  
\vspace{0.2cm}

\tabletypesize{\scriptsize}
\tablehead{
\colhead{J0735$-$6302} & \colhead{J1048$-$5838} & \colhead{J1226$-$3223} & \colhead{J1623$-$0841} & \colhead{J1739$-$2521} & \colhead{J1754$-$3014} & \colhead{J1839$-$0141} & \colhead{J1848$-$1243}\\
}
\startdata
18.37(7)(77.3$\sigma$)  & 1.9166(1)(12.5$\sigma$) & 1.2713(7)(4.5$\sigma$)   & 28.58(7)(34.8$\sigma$) &  0.3637(1)(12.5$\sigma$)  &  41.8(9)(20.5$\sigma$) & 0.68317(2)(33.3$\sigma$) & 6.892(4)(6.3$\sigma$)\\
18.88(7)(63.5$\sigma$)  & 0.88912(2)(11.75$\sigma$) & 2.550(3)(4.25$\sigma$)  &  14.57(2)(34.5$\sigma$)  &  0.3583(1)(12.3$\sigma$)  &  3.235(1)(18.8$\sigma$)  & 0.70325(2)(31.8$\sigma$) & 8.624(4)(6.0$\sigma$)\\
21.12(9)(55.0$\sigma$)  & 0.99261(2)(11.5$\sigma$) & 4.433(8)(3.8$\sigma$)   &  3.390(1)(33.8$\sigma$)   &  0.4429(1)(11.5$\sigma$)  &  16.38(1)(18.5$\sigma$) & 0.67232(2)(30.0$\sigma$) & 65.4(3)(6.0$\sigma$)\\
7.16(1)(53.8$\sigma$)   & 3.9436(5)(11.3$\sigma$) & 2.496(3)(3.8$\sigma$) &  3.258(1)(32.8$\sigma$)   &  0.4512(1)(11.5$\sigma$)  &  3.470(1)(18.3$\sigma$)  & 0.59866(2)(29.3$\sigma$) & 3.814(1)(6.0$\sigma$)\\
\enddata
\caption{Results of Lomb$-$Scargle tests on all eight RRATs. This table provides four periodicities (hours) for each RRATs in the pulse arrival time that are ranked in the order of peak significance. The errors of periodicities are in the first parentheses, and the significance in the second parentheses to compare with the 2.5$\sigma$ (roughly 99\% confidence) level. The error is calculated as random statistical error with 1$\sigma$
\label{tab3}}
\end{deluxetable*}

For PSR J1623$-$0841, the first, second, and third strongest power spectral density peaks of this RRAT are harmonics of each other, leading to a fundamental period of 3.39 hr. This period is consistent with our detection of only one on period for this RRAT in any observation. One possible cause of this periodicity that a single asteroid orbits the pulsar, producing emission, which is variable on the orbital timescale. This would also cause a perturbation in the TOAs for this RRAT. Assuming a circular orbit and a neutron star mass of 1.4 solar masses, we can calculate that the orbital radius of the asteroid would be roughly 4 $\times 10^{4}$ km, and using Equation 6 in \citet{cs08}, assuming an edge-on orbit, the upper limit of the asteroid mass would be 2 $\times 10^3$ $M_{\bigoplus}$ ($\sim$6.3 Jupiter mass) so that the perturbation in residuals is less than the 0.76 ms RMS. For the case of PSR J1839$-$0141, several peaks in the power spectrum have similar periods of roughly 0.68 hr, perhaps indicating a broad peak due to multiple asteroids or an asteroid belt. Using the average of the periodicities for the first three peaks, we can calculate an upper limit on the mass of a possible asteroid of 666 $M_{\bigoplus}$ ($\sim$2.1 Jupiter mass) in the same way to make the residuals within 1.80 ms RMS. Therefore, the existence of large mass asteroids is not precluded by the timing for these pulsars.

\section{Discussion}

\begin{figure}[!tp]
\centering 
\vspace*{-0.6cm}%
\includegraphics[trim=50 20 250 50, width=10.2cm]{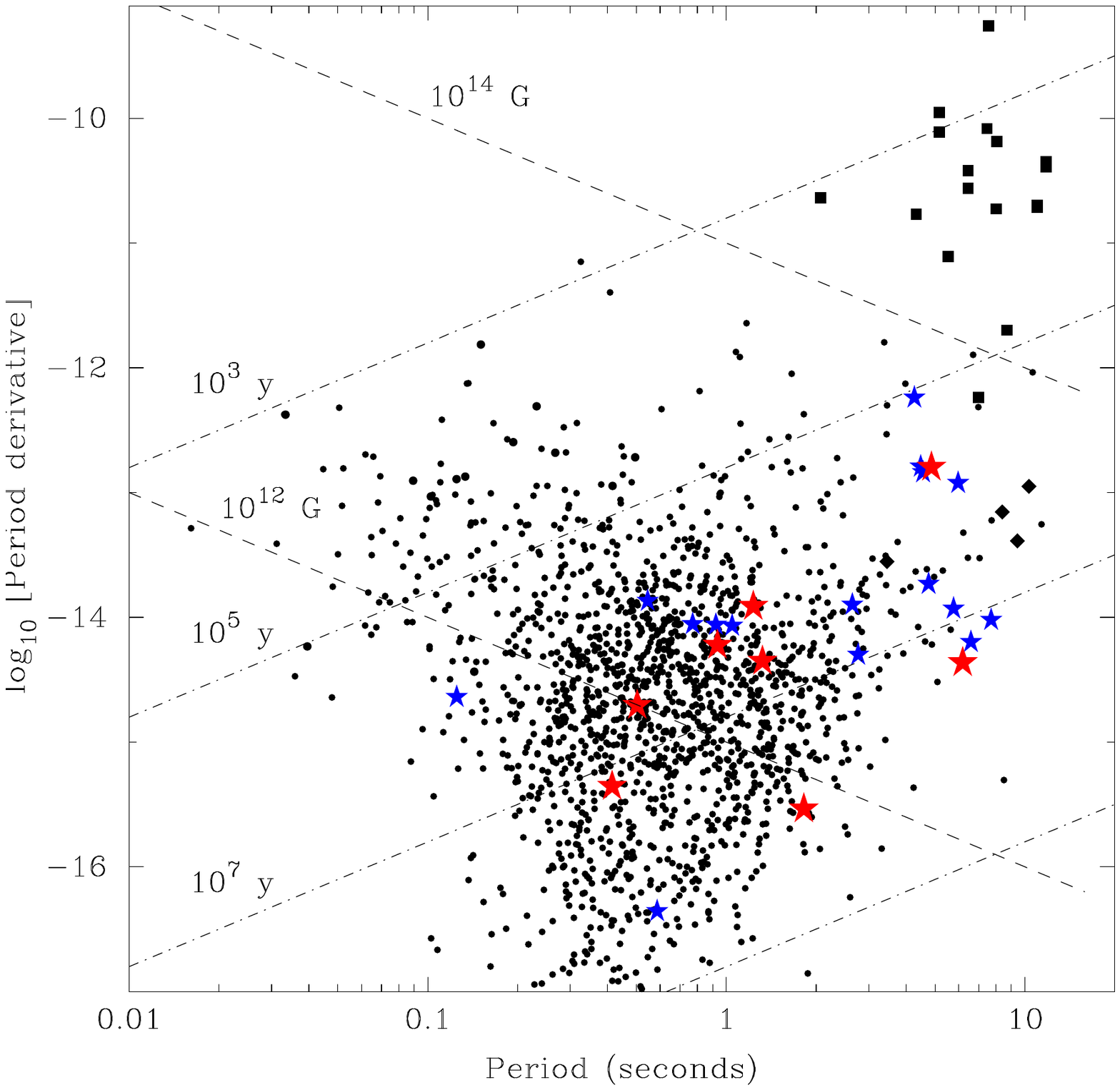}
\vspace*{-0.0cm}%
\caption{$P -\dot{P}$ diagram of RRATs and pulsars with timing solutions \citep{mhth05}. The RRATs with new timing solutions are shown as red stars and previously timed RRATs as blue stars. The black squares are magnetars with timing solutions \citep{crh+07}, and black diamonds indicate X-ray isolated neutron stars with timing solutions \citep{kk09}. Lines of constant magnetic field (dashed) and characteristic age (dotted-dashed) are shown. The KS test gives probabilities of 1.1$\times$10$^{-19}$, 2.5$\times$10$^{-4}$, 1.9$\times$10$^{-5}$, 0.16, and 0.04 that the period, period derivative, magnetic field, characteristic age, and spin-down energy-loss rate, respectively, were derived from the same distribution as those for other non-recycled pulsars.} 
\label{fig7}
\end{figure}

\begin{figure}[]
\vspace*{0.5cm}
\centering 
\includegraphics[width=9cm]{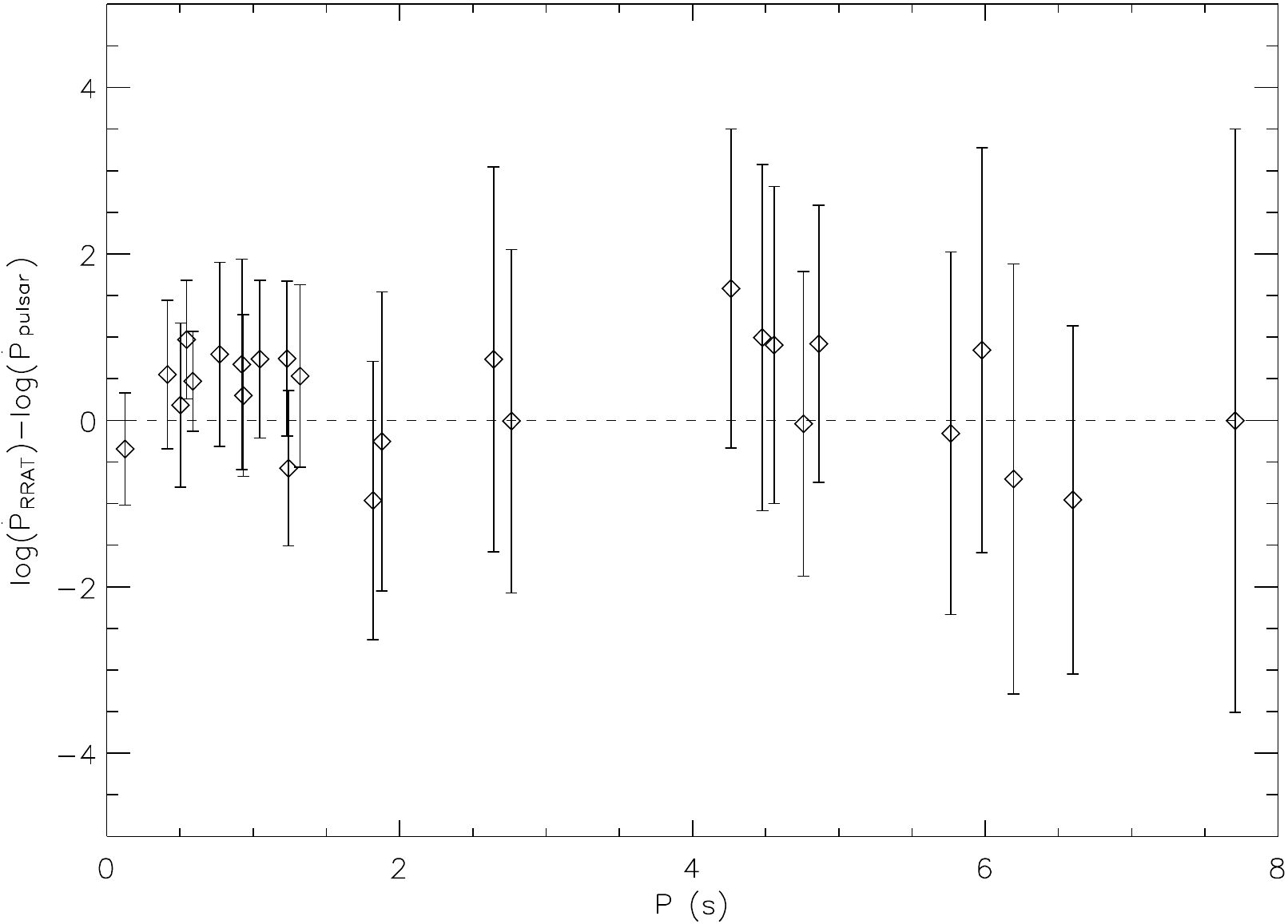}
\vspace*{0.1cm}
\caption{Comparison of period derivative between normal pulsars and RRATs. Each data point is the average value of $P$ and $\dot{P}$ of the population within a small period range. The Y-axis is the difference in logarithm values (\rm{log}($\dot{P}_{\rm{RRAT}})$ - \rm{log}($\dot{P}_{\rm{pulsar}}$)). We can see that for most period, the $\dot{P}$ values of the RRATs are larger than those of normal pulsars, so that the differences are above zero overall. All $P$ and $\dot{P}$ data used are from the ATNF Pulsar Catalog \citep{mhth05} and RRATalog website\footnote{see {\url{http://astro.phys.wvu.edu/rratalog}}}.}
\label{fig8}
\end{figure}

At this time, 25 of roughly 100 RRATs have timing solutions with period and period derivative, shown on the $P-\dot{P}$ diagram in Figure \ref{fig7}. We apply the Kolmogorov$-$Smirnov (KS) test to the RRAT and normal (non-recycled) pulsar populations to see how their spin-down properties compare. The result gives the probability $\mathscr{P}$ that two distributions are identical, and the largest differences between the two groups are found in the distributions of period (with $\mathscr{P}$ = 1.1$\times$10$^{-19}$) and magnetic field (with $\mathscr{P}$ = 1.9$\times$10$^{-5}$). While selection effects may be responsible for some of the period dependence, as longer period pulsars are more likely to be detected with higher S/Ns in single-pulse searches (see \citet{mc03} equation 2), the difference in period derivative (of $\mathscr{P}$=2.5$\times$10$^{-4}$), which along with the period is used to calculate the surface inferred magnetic field, hints that there is a fundamental difference in these populations. However, because young pulsars generally have higher period derivatives, period derivative is correlated with period, and it is therefore possible that the difference in period derivative distributions is due to a selection effect. We tested this by comparing the period derivatives of RRATs with pulsars with similar spin periods (see Figure \ref{fig8}). It is clear that the RRATs' period derivatives are higher than those of normal pulsars overall despite any selection effects.

We have provided timing solutions for eight RRATs, and also analyzed their pulse and emission properties. In these analyses, we find out that the amplitude distributions of RRATs generally follow a log-normal distribution, and some of the RRATs show periodicities in their emission timescales with a scenario of orbiting asteroids around these pulsars shown to be reasonable (other scenarios are still possible). We also explored how many pulses are missed in standard single-pulse searches and found that for some RRATs, this is a significant number. This analysis, as well as plots of the RRATs' weak pulse emission shows that some RRATs, like PSR J1048$-$5838, have truly sporadic emission, while most have a more continuous intensity distribution of which we detect only the brightest pulses.

However, to better understand their mechanism and evolution, we still need to extend our database of timing solutions and emission test results for more RRATs, in both quantity and quality. Further surveys for new RRATs, accompanied by sensitive timing observations and the development of new techniques, will help us achieve our goals.

\clearpage

\section*{Acknowledgments} 

We thank the West Virginia University for its financial support of GBT operations, which enabled some of the observations for this project. The National Radio Astronomy Observatory is a facility of the National Science Foundation operated under cooperative agreement by Associated Universities, Inc. The Parkes radio telescope is part of the Australia Telescope National Facility, which is funded by the Australian Government for operation as a National Facility managed by CSIRO.

\facility{Green Bank Telescope}, \facility{Parkes Telescope}.

\bibliographystyle{apj}
\bibliography{journals,modrefs,psrrefs}

\begin{thebibliography}{26}
\expandafter\ifx\csname natexlab\endcsname\relax\def\natexlab#1{#1}\fi

\bibitem[{{Boyles} {et~al.}(2013){Boyles}, {Lynch}, {Ransom}, {Stairs},
  {Lorimer}, {McLaughlin}, {Hessels}, {Kaspi}, \& {Kondratiev}}]{blr+13}
{Boyles}, J., {Lynch}, R.~S., {Ransom}, S.~M., {Stairs}, I.~H., {Lorimer},
  D.~R., {McLaughlin}, M.~A., {Hessels}, J.~W.~T., {Kaspi}, V.~M., \&
  {Kondratiev}, V.~I. 2013, \apj, 763, 80

\bibitem[{{Burke-Spolaor}(2013)}]{bur13}
{Burke-Spolaor}, S. 2013, Proceedings of the International Astronomical Union,
  291, 95

\bibitem[{{Burke-Spolaor} \& {Bailes}(2009)}]{bb09}
{Burke-Spolaor}, S., \& {Bailes}, M. 2009, \mnras, 1931

\bibitem[{{Burke-Spolaor} {et~al.}(2012){Burke-Spolaor}, {Johnston}, {Bailes},
  {Bates}, {Bhat}, {Burgay}, {Champion}, {D'Amico}, {Keith}, {Kramer}, {Levin},
  {Milia}, {Possenti}, {Stappers}, \& {van Straten}}]{bjb+12}
{Burke-Spolaor}, S., {Johnston}, S., {Bailes}, M., {Bates}, S.~D., {Bhat},
  N.~D.~R., {Burgay}, M., {Champion}, D.~J., {D'Amico}, N., {Keith}, M.~J.,
  {Kramer}, M., {Levin}, L., {Milia}, S., {Possenti}, A., {Stappers}, B.~W., \&
  {van Straten}, W. 2012, \mnras, 423, 1351

\bibitem[{{Camilo} {et~al.}(2007){Camilo}, {Ransom}, {Halpern}, \&
  {Reynolds}}]{crh+07}
{Camilo}, F., {Ransom}, S.~M., {Halpern}, J.~P., \& {Reynolds}, J. 2007, \apjl,
  666, L93

\bibitem[{{Cordes} \& {Lazio}(2002)}]{cl02}
{Cordes}, J.~M., \& {Lazio}, T.~J.~W. 2002, {astro-ph/0207156}

\bibitem[{{Cordes} \& {Shannon}(2008)}]{cs08}
{Cordes}, J.~M., \& {Shannon}, R.~M. 2008, \apj, 682, 1152

\bibitem[{{DuPlain} {et~al.}(2008){DuPlain}, {Ransom}, {Demorest}, {Brandt},
  {Ford}, \& {Shelton}}]{drd+08}
{DuPlain}, R., {Ransom}, S., {Demorest}, P., {Brandt}, P., {Ford}, J., \&
  {Shelton}, A. 2008, Advanced Software and Control for Astronomy II, Proc.
  SPIE 7019

\bibitem[{{Hobbs} {et~al.}(2006){Hobbs}, {Edwards}, \& {Manchester}}]{hem06}
{Hobbs}, G.~B., {Edwards}, R.~T., \& {Manchester}, R.~N. 2006, \mnras, 369, 655

\bibitem[{Jacoby {et~al.}(2009)Jacoby, Bailes, Ord, Knight, Hotan, Edwards, \&
  Kulkarni}]{jbo+09}
Jacoby, B.~A., Bailes, M., Ord, S., Knight, H., Hotan, A., Edwards, R.~T., \&
  Kulkarni, S.~R. 2009, \apj, 699, 2009

\bibitem[{{Kaplan} \& {van Kerkwijk}(2009)}]{kk09}
{Kaplan}, D.~L., \& {van Kerkwijk}, M.~H. 2009, \apjl, 692, L62

\bibitem[{{Keane} {et~al.}(2011){Keane}, {Kramer}, {Lyne}, {Stappers}, \&
  {McLaughlin}}]{kkl+11}
{Keane}, E.~F., {Kramer}, M., {Lyne}, A.~G., {Stappers}, B.~W., \&
  {McLaughlin}, M.~A. 2011, \mnras, 415, 3065

\bibitem[{{Keane} \& {McLaughlin}(2011)}]{km11}
{Keane}, E.~F., \& {McLaughlin}, M.~A. 2011, Bulletin of the Astronomical
  Society of India, 39, 333

\bibitem[{{Keane} \& {Petroff}(2015)}]{kp15}
{Keane}, E.~F., \& {Petroff}, E. 2015, \mnras, 447, 2852

\bibitem[{{Keith} {et~al.}(2009){Keith}, {Eatough}, {Lyne}, {Kramer},
  {Possenti}, {Camilo}, \& {Manchester}}]{kel+09}
{Keith}, M.~J., {Eatough}, R.~P., {Lyne}, A.~G., {Kramer}, M., {Possenti}, A.,
  {Camilo}, F., \& {Manchester}, R.~N. 2009, \mnras, 395, 837

\bibitem[{{Keith} {et~al.}(2010){Keith}, {Jameson}, {van Straten}, {Bailes},
  {Johnston}, {Kramer}, {Possenti}, {Bates}, {Bhat}, {Burgay}, {Burke-Spolaor},
  {D'Amico}, {Levin}, {McMahon}, {Milia}, \& {Stappers}}]{kjs+10}
{Keith}, M.~J., {Jameson}, A., {van Straten}, W., {Bailes}, M., {Johnston}, S.,
  {Kramer}, M., {Possenti}, A., {Bates}, S.~D., {Bhat}, N.~D.~R., {Burgay}, M.,
  {Burke-Spolaor}, S., {D'Amico}, N., {Levin}, L., {McMahon}, P.~L., {Milia},
  S., \& {Stappers}, B.~W. 2010, \mnras, 409, 619

\bibitem[{{Li}(2006)}]{li06}
{Li}, X.-D. 2006, \apjl, 646, L139

\bibitem[{{Luo} \& {Melrose}(2007)}]{lm07}
{Luo}, Q., \& {Melrose}, D. 2007, \mnras, 378, 1481

\bibitem[{{Manchester}(2001)}]{man01a}
{Manchester}, R.~N. 2001, PASA, 18, 1

\bibitem[{{Manchester} {et~al.}(2005){Manchester}, {Hobbs}, {Teoh}, \&
  {Hobbs}}]{mhth05}
{Manchester}, R.~N., {Hobbs}, G.~B., {Teoh}, A., \& {Hobbs}, M. 2005, Astron.\
  J., 129, 1993

\bibitem[{{McLaughlin} \& {Cordes}(2003)}]{mc03}
{McLaughlin}, M.~A., \& {Cordes}, J.~M. 2003, Astrophys.\ J., 596, 982

\bibitem[{{McLaughlin} {et~al.}(2006){McLaughlin}, {Lyne}, {Lorimer}, {Kramer},
  {Faulkner}, {Manchester}, {Cordes}, {Camilo}, {Possenti}, {Stairs}, {Hobbs},
  {D'Amico}, {Burgay}, \& {O'Brien}}]{mll+06}
{McLaughlin}, M.~A., {Lyne}, A.~G., {Lorimer}, D.~R., {Kramer}, M., {Faulkner},
  A.~J., {Manchester}, R.~N., {Cordes}, J.~M., {Camilo}, F., {Possenti}, A.,
  {Stairs}, I.~H., {Hobbs}, G., {D'Amico}, N., {Burgay}, M., \& {O'Brien},
  J.~T. 2006, \nat, 439, 817

\bibitem[{{Mickaliger} {et~al.}(2012){Mickaliger}, {McLaughlin}, {Lorimer},
  {Langston}, {Bilous}, {Kondratiev}, {Lyutikov}, {Ransom}, \&
  {Palliyaguru}}]{mml+12}
{Mickaliger}, M.~B., {McLaughlin}, M.~A., {Lorimer}, D.~R., {Langston}, G.~I.,
  {Bilous}, A.~V., {Kondratiev}, V.~I., {Lyutikov}, M., {Ransom}, S.~M., \&
  {Palliyaguru}, N. 2012, \apj, 760, 64

\bibitem[{{Palliyaguru} {et~al.}(2011){Palliyaguru}, {McLaughlin}, {Keane},
  {Kramer}, {Lyne}, {Lorimer}, {Manchester}, {Camilo}, \& {Stairs}}]{pmk+11}
{Palliyaguru}, N.~T., {McLaughlin}, M.~A., {Keane}, E.~F., {Kramer}, M.,
  {Lyne}, A.~G., {Lorimer}, D.~R., {Manchester}, R.~N., {Camilo}, F., \&
  {Stairs}, I.~H. 2011, \mnras, 417, 1871

\bibitem[{Scargle(1982)}]{sca82}
Scargle, J.~D. 1982, Astrophys.\ J., 263, 835

\bibitem[{{Weltevrede} {et~al.}(2006){Weltevrede}, {Stappers}, {Rankin}, \&
  {Wright}}]{wsrw06}
{Weltevrede}, P., {Stappers}, B.~W., {Rankin}, J.~M., \& {Wright}, G.~A.~E.
  2006, \apjl, 645, L149

\end{thebibliography}


\end{document}